\documentclass{JHEP3}
\pdfoutput=1

\usepackage{graphicx}
\usepackage[vcentermath]{youngtab}
\usepackage{amsmath,amsfonts}

\newcommand{\la}{\langle}
\newcommand{\ra}{\rangle}

\newcommand{\braket}[1]{\mathinner{\left\langle{#1}\right\rangle}}

\newcommand{\sy}{\text{sym}}
\newcommand{\as}{\text{asym}}
\newcommand{\tr}{\text{true}}
\newcommand{\SU}{\text{SU}}
\newcommand{\1}{\mathbf{1}}
\newcommand{\Z}{\mathbb{Z}}

\DeclareMathOperator{\re}{Re}
\DeclareMathOperator{\im}{Im}
\DeclareMathOperator{\diag}{diag}
\DeclareMathOperator{\Tr}{Tr}

\title{Eigenvalue density of Wilson loops in 2D SU(N) YM}
\author{Robert Lohmayer,$^a$ Herbert Neuberger,$^b$ and Tilo
  Wettig$^a$ \\ 
  \llap{$^a$}Institute for Theoretical Physics, University of
  Regensburg, 93040 Regensburg, Germany\\ 
  \llap{$^b$}Department of Physics and Astronomy, Rutgers University, 
  Piscataway, NJ 08855, USA\\
  Email: \email{robert.lohmayer@physik.uni-regensburg.de},
  \email{neuberg@physics.rutgers.edu},
  \email{tilo.wettig@physik.uni-regensburg.de}} 

\abstract {In 1981 Durhuus and Olesen (DO) showed that at infinite $N$
  the eigenvalue density of a Wilson loop matrix $W$ associated with a
  simple loop in two-dimensional Euclidean $\SU(N)$ Yang-Mills theory
  undergoes a phase transition at a critical size. The averages of
  $\det(z-W)$, $\det(z-W)^{-1}$, and $\det(1+uW)/(1-vW)$ at finite $N$
  lead to three different smoothed out expressions, all tending to the
  DO singular result at infinite $N$.  These smooth extensions are
  obtained and compared to each other.}

\keywords{Large N, Lattice Gauge Field Theories}

\preprint{April 27, 2009}

\begin{document}

\section{Introduction} 

So far, the large-$N$ limit of $\SU(N)$ YM theory has been mainly
employed for qualitative insights. Our objective is to find a way to
exploit the simplifications at large $N$ for actual first-principles
quantitative calculations of at least some nonperturbative quantities
in theories that interact weakly at short distances and strongly at
long distances. To be sure, significantly more work is needed in order
to construct a real calculational framework; we are not there yet.

In two Euclidean dimensions the eigenvalue distribution of the
$\SU(N)$ Wilson matrix associated with a non-selfintersecting loop
undergoes a phase transition in the infinite-$N$ limit as the loop is
dilated \cite{Durhuus:1980nb}.  This phase transition has universal
properties shared across dimensions and across analog two-dimensional
models \cite{Narayanan:2007dv,Narayanan:2008he}. Thus, a detailed
understanding of the transition region in 2D is of relevance to
crossovers from weakly to strongly interacting regimes in a wide class
of models based on doubly-indexed dynamical variables with symmetry
$\SU(N)$.  Building upon previous work
\cite{Neuberger:2008mk,Neuberger:2008ti,Blaizot:2008nc}, this paper
presents several new results in this context and indicates how such
results might be used to estimate long-distance parameters by
analytical means, at least for $N\gg1$.

We are focusing on the eigenvalues of the Wilson loop. The associated
observables are three different functions $\rho^\ell_N (\theta)$, with
$\ell=\as$, $\sy$, $\tr$, of an angular variable $\theta$. At infinite
$N$ the three functions have identical limits: $\rho^\ell_\infty
(\theta)=\rho_\infty (\theta )$. For a specific critical scale, the
nonnegative function $\rho_\infty (\theta )$ exhibits a transition at
which a gap centered at $\theta=\pm \pi$, present for small loops,
just closes. This transition was discovered by Durhuus and Olesen in
1981~\cite{Durhuus:1980nb}.

For the time being, we shall suppress the size dependence of the
$\rho^\ell_N (\theta)$.  $\rho^\ell_N (\theta)$ for $\ell=\as$, $\sy$
are obtained from the logarithmic derivative of $\langle \det^k
(z-W)\rangle$ for $k=1$ and $-1$, respectively.  One needs to take $z$
to $e^{i\theta}$ in a specified manner. Neither of these two functions
$\rho^\ell_N (\theta)$ has a natural interpretation at finite $N$; the
interest in these functions mainly stems from them obeying simple
partial integro/differential equations which are exactly integrable
and already known and studied in other
contexts~\cite{Neuberger:2008mk,Neuberger:2008ti}.  In the course of
this paper we extend the results
in~\cite{Neuberger:2008mk,Neuberger:2008ti}.

Unlike $\rho^\ell_N (\theta)$ with $\ell=\as$, $\sy$, $\rho^{\tr}_N
(\theta)$ literally is the eigenvalue density at finite $N$, $\langle
\Tr \delta( \theta +i\log W )\rangle$, and poses no difficulties of
interpretation.  It can be obtained from $\langle
\det(1+uW)/(1-vW)\rangle$ in the limit $-u\to v\to e^{i\theta}$.  We
shall derive expressions for $\rho^{\tr}_N (\theta)$ in this work.  As
anticipated in~\cite{Neuberger:2008ti}, we find no evidence that
$\rho^{\tr}_N (\theta)$ obeys as simple equations as the $\rho^\ell_N
(\theta)$ for $\ell\ne \tr$ were found to do.

This paper starts with a general description of $\rho^\ell_N
(\theta)$.  We then follow with details for each case. First, we
describe the case $\ell=\as$, where the focus is on the loop-size
dependence of the zeros $z_j$ of the average characteristic
polynomial. The equations governing these zeros were derived
in~\cite{Neuberger:2008mk}, and here we work out the approximate
solution for small, intermediate, and large loops.  Then comes a
description of the case $\ell=\sy$ and a saddle-point analysis of the
integral representation found in~\cite{Neuberger:2008ti}. A connection
to the multiplicative random matrix model
of~\cite{GudowskaNowak:2003zx,Lohmayer:2008bd} is pointed out.  We
then proceed to deriving exact representations of
$\rho^{\tr}_N(\theta)$.  As anticipated, we do not find a simple
direct equation for $\rho^{\tr}_N(\theta)$, but we do find a simple
equation for $\langle \det(1+uW)/(1-vW)\rangle$.  From this we obtain
a representation of $\rho^{\tr}_N(\theta)$ by a sum.  By numerically
performing the sum, $\rho^{\tr}_N(\theta)$ can be evaluated to any
desired accuracy. Further, we obtain an integral representation for
$\rho^{\tr}_N(\theta)$ which is useful for setting up the $1/N$
expansion of $\rho^{\tr}_N(\theta)$.  We carry out the saddle-point
analysis which is the starting point of this expansion. We also show
that one can define a natural extension to negative values of $N$, and
in this extension $\rho^{\tr}_N(\theta)=\rho^{\tr}_{-N}(\theta)$.  We
follow this by a numerically aided study of the relations between the
three $\rho^\ell_N(\theta)$.  We compare numerically the densities
$\rho^{\tr}_N (\theta)$ and $\rho^{\sy}_N (\theta )$ at the same
areas.  As $\rho^{\as}_N (\theta)$ is given by a sum of
$\delta$-functions, its comparison to another $\rho^\ell_N (\theta)$
is less direct.  We conjecture, and check numerically, that the
location of the $N$ peaks in $\rho^{\tr}_N(\theta)$ are close to the
matching zeros $\theta_j=-i\log z_j$ of the average characteristic
polynomial. By ``close'' we mean that for large $N$ the distance
between a $\theta_j$ and the matching peak vanishes faster than the
distance between that peak and its adjacent valley.

In order to indicate how this paper fits into a larger research plan,
we finish with a sketch of the bigger motivating picture.

\section{\boldmath Three ``densities'' $\rho^\ell_N (\theta )$ and how
  they compare}

\subsection{Different definitions of dimensionless area}

The dimensionless area variable has to take a slightly different form
for the average of the characteristic polynomial and the average of
its inverse to obey equations that look simple.

We define
\begin{equation}
  t={\cal A} g^2 N\,,
\end{equation}
where ${\cal A}$ is the area enclosed by the Wilson loop, $g$ is the
YM coupling, and the gauge group is $\SU(N)$. $\lambda=g^2 N$ is the
standard 't Hooft coupling, and $t$ makes it dimensionless. This $t$
appears in $\rho^{\tr}_N (\theta, t)$.

The average characteristic polynomial generates the expectation values
of the characters of all antisymmetric representations of the Wilson
loop matrix. The appropriate area variable in this case is denoted by
$\tau$, with \cite{Neuberger:2008ti}
\begin{equation}
  \label{defTau}
  \tau=t\left(1+\frac1N\right)\,.
\end{equation}
Thus, when $\rho^{\as}_N (\theta,\tau)$ is compared to $\rho^{\tr}_N
(\theta, t)$, the $1/N$ correction in $t$ relative to $\tau$ has to be
taken into account.

The average of the inverse of the characteristic polynomial generates
the expectation values of the characters of all symmetric
representations of the Wilson loop matrix. The appropriate area
variable in this case is denoted by $T$, with \cite{Neuberger:2008ti}
\begin{equation}
  \label{defT}
  T=t\left(1-\frac1N\right)\,.
\end{equation}
Thus, when $\rho^{\sy}_N (\theta,T)$ is compared to $\rho^{\tr}_N
(\theta, t)$, the $1/N$ correction in $t$ relative to $T$ has to be
taken into account.

\subsection[Averaging over the $\SU(N)$ Wilson loop matrix
$W$]{\boldmath Averaging over the $\SU(N)$ Wilson loop matrix $W$}

The probability density for $W$ is given by the heat kernel (see for
example~\cite{Gross:1993hu} and original references therein)
\begin{equation}
  \label{eq:weight}
  {\cal P}_N (W,t) = \sum_r d_r \chi_r (W) e^{-\frac{t}{2N} C_2 (r)}
\end{equation}
with $t=\lambda {\cal A}$, where the sum over $r$ is over all distinct
irreducible representations of $\SU(N)$ with $d_r$ denoting the
dimension of $r$ and $C_2 (r)$ denoting the value of the quadratic
Casimir on $r$. $\chi_r (W)$ is the character of $W$ in the
representation $r$ and is normalized by $\chi_r (\1) = d_r$.  Averages
over $W$ at fixed $t$ are given by
\begin{equation}
  \langle {\cal O}(W)\rangle = \int dW {\cal P}_N (W,t )\mathcal{O}(W)\,,
\end{equation}
where $dW$ is the Haar measure on $\SU(N)$ normalized by $\int dW =1$.
Note that we have $\int dW {\cal P}_N(W,t)=1$.
Any class function can be averaged when expanded in characters using
character orthogonality.  

Because in the sum over $r$ in \eqref{eq:weight} each representation
is accompanied by its complex conjugate representation, it is easy to
see that
\begin{equation} 
  \langle {\cal O}(W)\rangle = \langle {\cal O}(W^\dagger)\rangle =
  \langle {\cal O}(W^\ast)\rangle\, , 
\label{symmetry}
\end{equation}
implying identities relating $\langle \det(z-W)\rangle$, $\langle
\det(z-W)^{-1}\rangle$, and $\langle \det(1+uW)/(1-vW)\rangle$ to the
same objects with $z\to 1/z$, $z\to z^*$, $u,v\to 1/u ,1/v$, and
$u,v\to u^*, v^*$, respectively.

\subsection[General features of the $\rho^\ell_N(\theta)$]{\boldmath General features of the $\rho^\ell_N(\theta)$}

The $\rho^\ell_N$ are real on the unit circle parametrized by the
angle $|\theta|\le\pi$, even under $\theta\to-\theta$, and depend on
the size of the loop. All three are positive distributions in
$\theta$, normalized by
\begin{equation}
  \int_{-\pi}^\pi \frac{d\theta}{2\pi} \rho^\ell_N (\theta )=1\,.
\end{equation}
$\rho^{\as}_N$ summarizes the averages of the characters of $W$ in all
totally antisymmetric representations, i.e., single-column Young
diagrams.  $\rho^{\sy}_N$ summarizes the averages of the characters of
$W$ in all totally symmetric representations, i.e., single-row Young
diagrams.  $\rho^{\tr}_N$ summarizes the averages of the traces of all
$k$-wound Wilson loops matrices, $\langle \Tr W^k \rangle$.  As we
will discuss in section~\ref{sec:true}, the latter are determined by
linear combinations of the averages of the characters of $W$ in
representations which we label by $(p,q)$ and whose Young diagrams
have the following shape:
\begin{equation}
  \label{youngd}
  \young(\hfil12\hfil\hfil q,1,2,\hfil,\hfil,p)
\end{equation}
$\rho^{\tr}_N$ determines $\langle \Tr f(W)\rangle $ for any function
$f$.  However, unlike $\rho^{\as}_N$ and $\rho^{\sy}_N$, it has no
information about any average of the type $\langle \Tr f(W) \Tr g(W)
\rangle$, where the number of trace factors exceeds one. In other
words, $\rho^{\tr}_N$ is the single eigenvalue density and, unlike
$\rho^{\as}_N$ and $\rho^{\sy}_N$, contains no information about any
higher-point eigenvalue correlations.

\subsection[$\rho^{\as}_N (\theta,\tau)$]{\boldmath $\rho^{\as}_N
  (\theta,\tau)$}
\label{subsecRhoAsym}

$\rho^{\as}_N (\theta ,\tau)$ is constructed from the logarithmic
derivative of the average of the characteristic polynomial, whose
zeros are $z_j(\tau)=\exp(i\theta_j(\tau))$ with $j=0,\ldots,N-1$ and
$-\pi\le \theta_j \le \pi$,
\begin{equation}
  \label{eq:det}
  \psi^{(N)}(z,\tau)\equiv\langle\det(z-W)\rangle= \prod_{j=0}^{N-1}
  (z - z_j (\tau) )\,. 
\end{equation}
Define
\begin{align}
  \phi^{(N)} (z,\tau)=\frac{i}{N} \frac{1}{\psi^{(N)} (z,\tau)} \left
    [ z\frac{\partial}{\partial z} +\frac{N}{2}\right ] \psi^{(N)}
  (z,\tau)\,.
\end{align}
Setting $z=e^{-iy}$ we obtain
\begin{equation}
  \phi^{(N)} (e^{-iy},\tau) = i-\frac{1}{N}\sum_{j=0}^{N-1}
  \sum_{n\in \Z}\frac{1}{y+\theta_j + 2\pi n}\,.
\end{equation}
We set $y$ to $\theta\pm i \epsilon$ with $\epsilon >0$ and real
$\theta$ and define
\begin{equation}
  \varphi^{(N)}_\pm (\theta,\tau)=\lim_{\epsilon\to0} \phi^{(N)}
  (e^{-i(\theta\pm i\epsilon)},\tau )\,.
\end{equation}
Finally, we define $\rho^{\as}_N (\theta,\tau)$ in analogy to
$\rho^{\sy}_N (\theta,T)$ in~\cite{Neuberger:2008ti},
\begin{align}
  \rho^{\as}_N (\theta,\tau) &= -2\re \left[i\varphi^{(N)}_+
    (\theta,\tau)+1\right] =-i\left [ \varphi^{(N)}_+ (\theta,\tau ) -
    \varphi^{(N)}_-
    (\theta,\tau)\right ]\notag\\
  &=\frac{2\pi}{N} \sum_{j=0}^{N-1} \delta_{2\pi} \left
    (\theta+\theta_j (\tau) \right ) =\frac{2\pi}{N} \sum_{j=0}^{N-1}
  \delta_{2\pi} \left (\theta-\theta_j (\tau) \right )\,,
  \label{eq:rhoas}
\end{align}
where $\delta_{2\pi}$ denotes the $2\pi$-periodized $\delta$-function
with normalization
\begin{align}
  \int_{-\pi}^\pi d\theta \, \delta_{2\pi}(\theta)=1
\end{align}
and the last identity in \eqref{eq:rhoas} follows from the fact that
$\rho^{\as}_N (\theta,\tau)$ is even in $\theta$.  The sum over
$\delta$-functions will reproduce exactly the averages of the traces
of $W$ in all totally antisymmetric representations at arbitrary
finite $N$, simply by setting $W$ equal to $\diag(e^{i\theta_0
  (\tau)}, e^{i\theta_1 (\tau)},\ldots ,e^{i\theta_{N-1}
  (\tau)})$. Thus, the entire information of $\rho^{\as}_N (\theta
,\tau)$ is contained in the set $\theta_j (\tau)$.  It is obvious that
given $\rho^{\as}_N (\theta, \tau)$ we can reconstruct
$\phi^{(N)}(z,\tau)$ and $\psi^{(N)}(z,\tau)$.  The infinite-$N$ limit
of $\rho^{\as}_N (\theta, \tau)$ is $\rho_\infty(\theta, \tau)$
\cite{Narayanan:2007dv}.

In~\cite{Neuberger:2008mk} it was shown that the $\theta_j(\tau)$ are
determined by a set of first-order ``equations of motion'' in $\tau$
with a specific initial condition,
\begin{equation}
  \dot \theta_j\equiv\frac{\partial\theta_j}{\partial\tau} 
  =\frac{1}{2N}\sum_{k\ne j} \cot\frac{\theta_j-\theta_k}{2}\,.
\label{eqmm}
\end{equation}
The initial condition 
\begin{equation}
  \theta_j(0)=0
\end{equation}
is at a singular point of the differential equations. However, once
one understands that as $\tau$ grows from zero the $\theta_j(\tau)$
spread out, the solution becomes uniquely determined. Throughout the
evolution, the $\dot\theta_j$ never change sign. For any $\tau >0$ we
have
\begin{equation}
  \theta_0 (\tau) < \theta_1 (\tau) <\ldots < \theta_{N-1} (\tau)\,.
\end{equation}
There is a $\Z_2$ symmetry pairing them,
\begin{equation}
  \theta_{N-j-1} (\tau)= -\theta_j (\tau)\,.
  \label{esign}
\end{equation}
If $N$ is odd~(\ref{esign}) yields
\begin{equation}
  \theta_{\frac{N-1}{2}}(\tau)= 0\,.
\end{equation}
Thus, there are $\left [ N/2 \right ] $ pairs of nonzero eigenvalues
of opposite signs, implying $\rho^{\as}_N(\theta,\tau)=\rho^{\as}_N
(-\theta,\tau)$.

In section~\ref{sec:asym} we shall calculate the behavior of the
$\theta_j (\tau)$ at small, critical, and large $\tau$.

\subsection[$\rho^{\sy}_N (\theta, T)$]{\boldmath $\rho^{\sy}_N
  (\theta, T)$}
\label{subsecRhoSymm}

$\rho^{\sy}_N (\theta ,T)$ is constructed from the logarithmic
derivative of the average of the inverse characteristic polynomial. We
reproduce here the relevant formulas from~\cite{Neuberger:2008ti}.
Define
\begin{equation}
  \psi^{(N)}_\pm (z, T)=\langle\det(z-W)^{-1}\rangle\,,
\end{equation}
where $+$ is for $|z|>1$ and $-$ for $|z|<1$. Because of the negative
power, one cannot exclude singularities at $|z|=1$ (although
equation~(22) of \cite{Neuberger:2008ti} shows that these
singularities are removable so that $\psi_\pm^{(N)}(z,T)$ can be
continued to $|z|=1$).  One should think about $\psi^{(N)}_\pm (z,T)$
as two distinct functions.  They are simply related to each other by
\begin{equation}
  \label{eq:psi-}
  \psi^{(N)}_- \left (1/z, T \right )=(-z)^{N} \psi^{(N)}_+ (z,T)\,,\qquad
  |z|>1\,.
\end{equation}
We now define
\begin{equation}
  \label{eq:phipm}
  \phi^{(N)}_\pm (z,T) =\frac{i}{N}\frac{1}{\psi^{(N)}_\pm (z,T)}
  \left ( z\frac{\partial}{\partial z} +\frac{N}{2}\right )
  \psi^{(N)}_\pm (z,T)\,. 
\end{equation}
$\rho^{\sy}_N (\theta ,T)$ is given by
\begin{equation}
  \label{eq:rhosym}
  \rho^{\sy}_N (\theta ,T)=i \lim_{\epsilon\to0}
  \left[\phi^{(N)}_+ (e^{-i\theta+\epsilon},T)
    -\phi^{(N)}_- (e^{-i\theta-\epsilon},T) \right]\,.
\end{equation}
Unlike $\rho^{\as}_N(\theta,\tau)$, $\rho^{\sy}_N (\theta ,T)$ is a
smooth function of $\theta$ for any finite $N$ and $T>0$.  It again
obeys $\rho^{\sy}_N (\theta ,T)= \rho^{\sy}_N (-\theta ,T)$. The
function is monotonic on each of the segments $(-\pi,0)$ and $(0,\pi)$
with the maximum at $\theta=0$ and the minimum at $\theta=\pm\pi$.
The infinite-$N$ critical point is at $T=4$. For $T>4$, $\rho^{\sy}_N
(\theta ,T)$ approaches $\rho_\infty (\theta,T)$ by power corrections
in $1/N$ \cite{Neuberger:2008ti}.  For $T<4$, $\rho_\infty (\theta,T)$
is zero for $|\theta|>\theta_c(T)$, where $0<\theta_c(T)<\pi$ and
$\theta_c(4)=\pi$.  In this interval $\rho^{\sy}_N (\theta ,T)$
approaches zero by corrections that are exponentially suppressed in
$N$.  $\rho^{\sy}_N (\theta ,T)$ has an explicit form in terms of
rapidly converging infinite sums,
\begin{align}
  \rho^{\sy}_N (\theta, T)&= 1+\frac{p(\theta,T)+p^*(\theta,T)}{N}\,,\\
  p(\theta, T)&=\frac{\sum_{k=1}^\infty k {N+k-1\choose N-1}
    e^{ik\theta} e^{-T\frac{k(k+N)}{2N}}}{1+\sum_{k=1}^\infty
    {N+k-1\choose N-1} e^{ik\theta} e^{-T\frac{k(k+N)}{2N}}}\,.
  \label{rhosym}
\end{align}
Given $\rho^{\sy}_N (\theta, T)$ with $T>0$ we can reconstruct
$\phi^{(N)}_\pm (z,T)$ and $\psi^{(N)}_\pm (z,T)$ using the Poisson
integral, on account of the analyticity of $\psi^{(N)}_- (z,T)$ for
$|z|<1$.

In~\cite{Neuberger:2008ti} it was also shown that $\psi^{(N)}_+ (z,T)$
for $|z|>1$ has an integral representation given by
\begin{equation}
  \psi^{(N)}_+ (z,T) = e^{\frac{NT}{8} } \sqrt{\frac{N}{2\pi T}}
  \int_{-\infty}^\infty du \, e^{-\frac{N}{2T}u^2} \left ( z
    e^{-i\frac{u}{2}} - e^{i\frac{u}{2}}\right )^{-N}\,. 
  \label{psint}
\end{equation}
It was pointed out there that this formula exhibited a formal relation
to $\langle\det(z-W)\rangle$ under a sign switch of $N$. Similar
observations have been made in the past, see~\cite{Dunne:1988ih} and
references therein.

Equation \eqref{rhosym} can be evaluated numerically for arbitrary $N$
to any desired precision.  In figure \ref{figRhoSymm} we show how
$\rho_N^\sy(\theta,T)$ approaches the infinite-$N$ result
$\rho_\infty(\theta)$ of DO~\cite{Durhuus:1980nb} for fixed $T=2$ and
$T=5$. %
\FIGURE[t]{
  \includegraphics[width=0.43\textwidth]{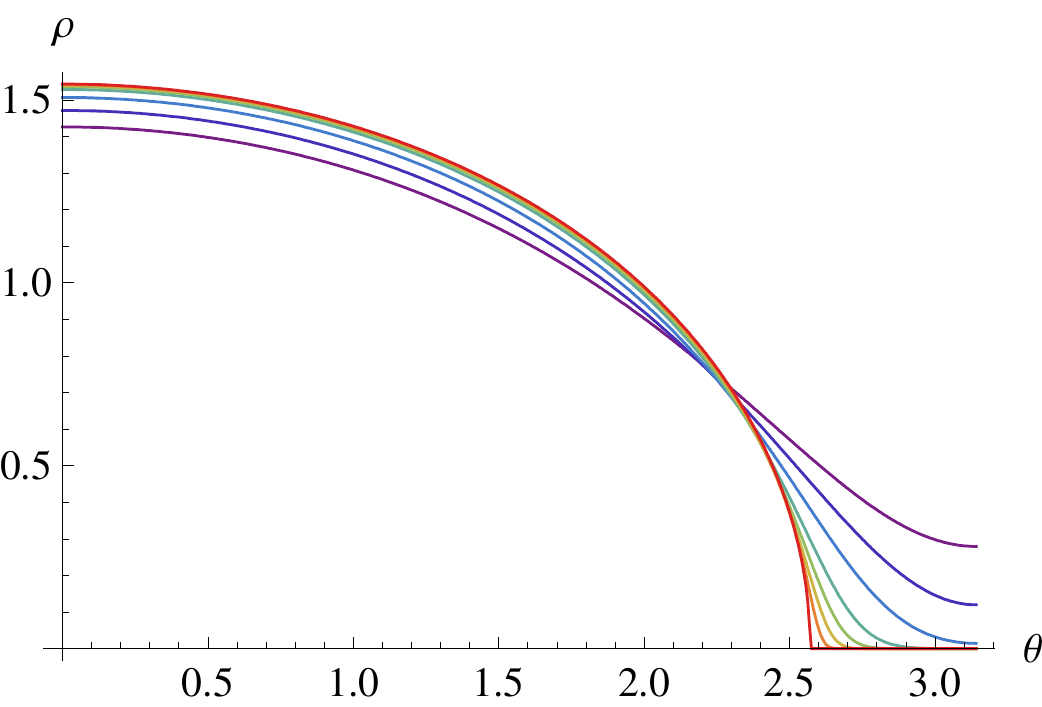}\hfill    
  \includegraphics[width=0.43\textwidth]{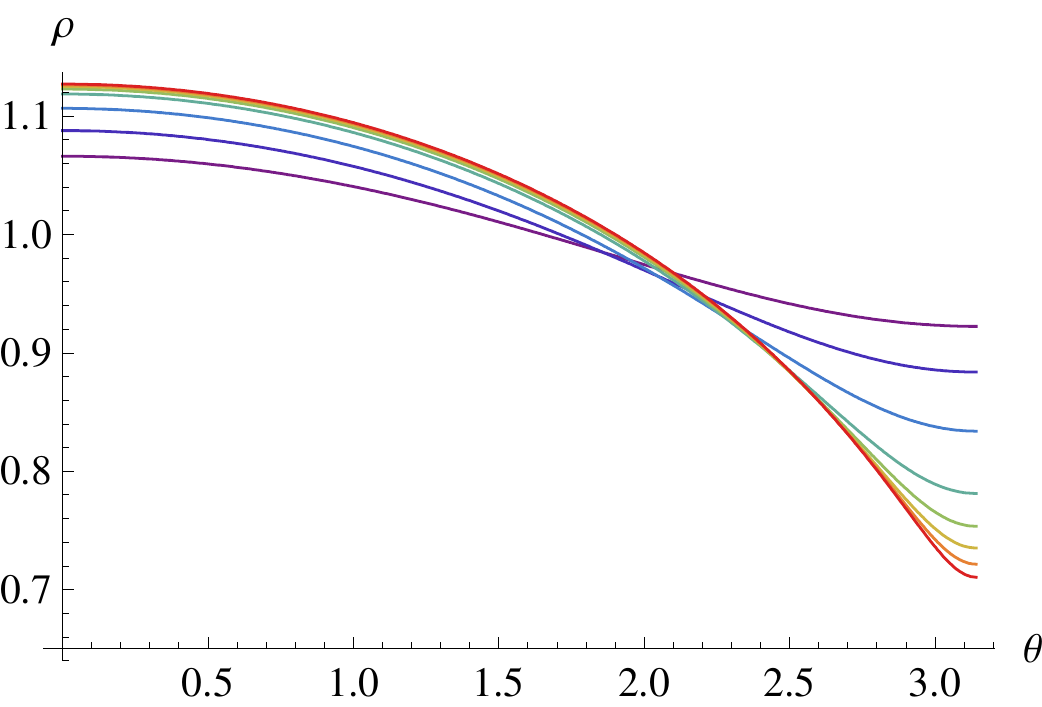}\hfill
  \vspace{-2cm}    
  \includegraphics[width=0.09\textwidth]{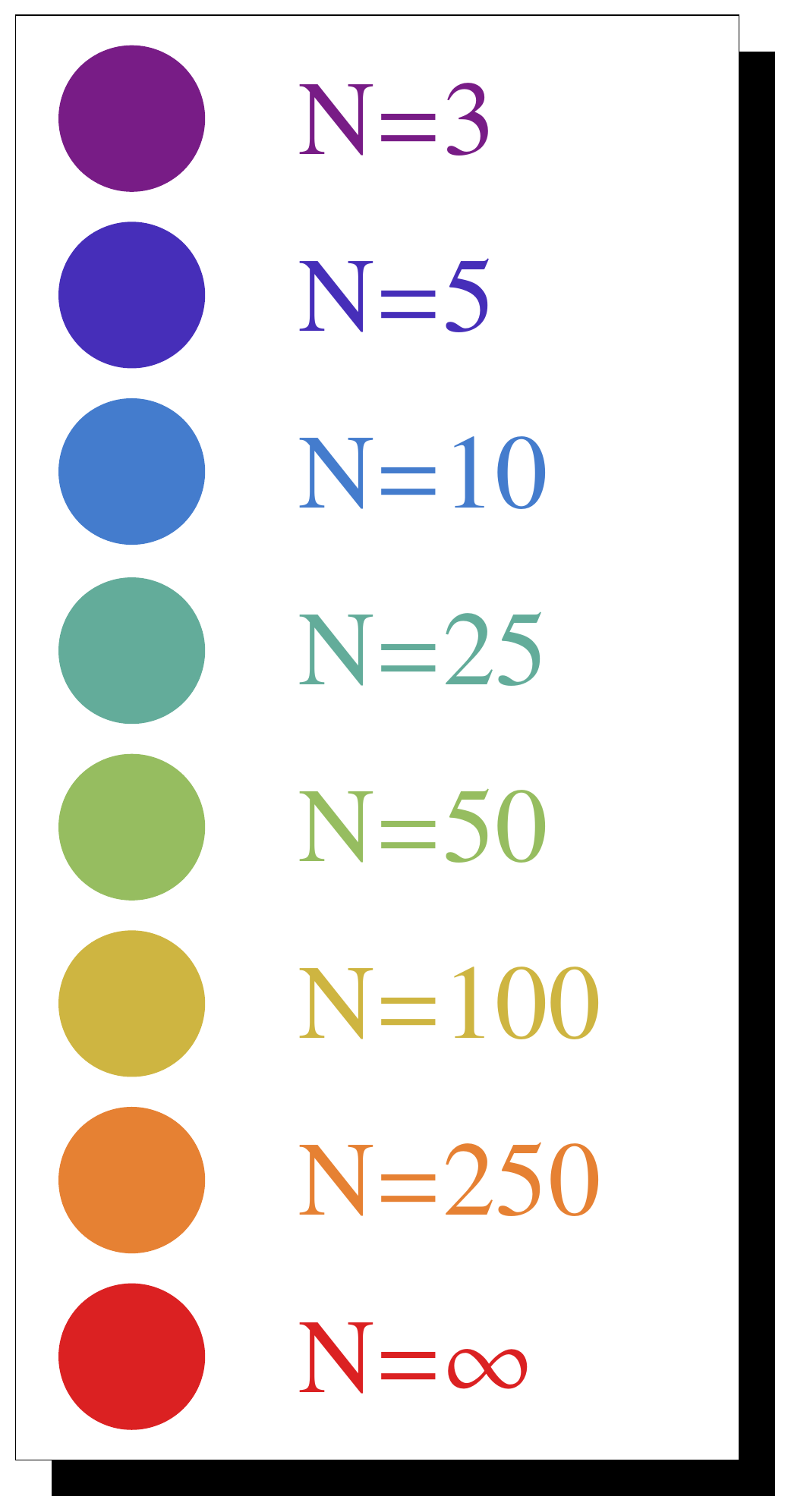} 
  \vspace{2cm}
  \caption{Plots of $\rho_N^\sy(\theta,T)$ for $T=2$ (left), $T=5$
    (right), and $N=3,5,10,25,50,100,250$ together with
    $\rho_\infty(\theta,T)$.}
  \label{figRhoSymm}
} %
In addition to these numerical results, it would be useful to compute
analytically the asymptotic expansion of $\rho^{\sy}_N (\theta, T)$ in
$1/N$.  For this it is enough to expand $\psi^{(N)}_+ (z,T)$ in $1/N$,
which is best done by starting from~(\ref{psint}).  The $1/N$
expansion then comes from an expansion around a single saddle point.
This problem will be considered in section~\ref{secSymmSaddle}.  The
saddle points turn out to be related to the position of the boundary
of the eigenvalue domain of the random multiplicative complex matrix
ensemble studied in~\cite{GudowskaNowak:2003zx,Lohmayer:2008bd}.

In section~\ref{sec:compare} we will show plots comparing
$\rho^{\sy}_N (\theta, T)$ to $\rho^{\tr}_N (\theta, t)$.

\subsection[$\rho^{\tr}_N (\theta,t)$]{\boldmath $\rho^{\tr}_N
  (\theta,t)$}
\label{sec:tr}

Finally, unlike $\rho^\ell_N$ with $\ell=\as,\sy$,
$\rho^{\tr}_N(\theta,t)$ has a natural definition.  If the eigenvalues
of $W$ are $e^{i\alpha_j}$ with $j=0,1,\ldots,N-1$, we define
\begin{align}
  \rho^{\tr}_N(\theta,t)
  =\frac{2\pi}{N}\sum_j\langle \delta_{2\pi}(\theta-\alpha_j(W))\rangle
  =\frac{2\pi}{N}\langle \Tr\delta_{2\pi}(\theta+i\log(W))\rangle\,.
\end{align}
With the help of $\rho^{\tr}_N(\theta,t)$ we can compute the averages
of a specific subset of class functions $F(W)$, namely, those that can
be written as
\begin{equation}
  F(W)=\frac1N \sum_j f(\alpha_j (W))\,.
\end{equation}
The obvious formula is
\begin{equation}
  \langle F(W)\rangle 
  = \Bigl\langle \frac{1}{N} \sum_j \int_{-\pi}^\pi
  d\theta f(\theta) \delta_{2\pi} (\theta-\alpha_j (W)) \Bigr \rangle
  =\int_{-\pi}^\pi\frac{d\theta}{2\pi} f(\theta) \rho^{\tr}_N(\theta,t)\,. 
\end{equation}
$\rho^{\tr}_N(\theta,t)$ summarizes all the information contained in
the entire collection of averages of the type $\langle \Tr
f(W)\rangle$.  Viewed in this way, it is analogous to
$\rho^{\as}(\theta,\tau)$ and $\rho^{\sy}(\theta,T)$, which summarize
all the information contained in all averages $\langle \chi_r (W)
\rangle $, with $r$ denoting all totally antisymmetric and all totally
symmetric representations, respectively.  The analog of the functions
$\phi^{(N)}(z,\tau)$ and $\phi^{(N)}_\pm (z, T)$ related to
$\rho^{\as}(\theta,\tau)$ and $\rho^{\sy}(\theta,T)$, respectively, in
the present case is the average resolvent,
\begin{equation}
  \label{eq:tr}
  \Phi^{(N)}_\pm (z,t) =\frac{1}{N} \braket{ \Tr\frac{1}{z-W} }\,.
\end{equation}
Here again the $+$ sign goes with $|z| >1 $ and the $-$ sign goes with
$|z|<1$.

Using~(\ref{symmetry}) one easily concludes that $\Phi^{(N)}_+ (z,t)$
determines $\Phi^{(N)}_- (z,t)$ just as in the case of $\phi^{(N)}_\pm
(z,T)$. Clearly, $\rho^{\tr}_N(\theta, t)$ determines $\Phi^{(N)}_\pm
(z, t)$ since the latter is the average of a single trace. It is easy
to see that the opposite is true also, namely, $\Phi^{(N)}_\pm (z,t)$
determines $\rho^{\tr}_N(\theta, t)$. If we use the restrictions
following from~(\ref{symmetry}), it is enough to use just
$\Phi^{(N)}_+ (z,t)$ for example,
\begin{equation}
  \label{eq:true}
  \rho^{\tr}_N(\theta, t) =2 \lim_{\epsilon\to 0^+} \re \left [
    e^{i\theta+\epsilon} \Phi_+^{(N)} (e^{i\theta +\epsilon}, t)\right ] -1\,.
\end{equation}
$\rho^{\tr}_N(\theta,t)$ is smooth over the circle and similar to
$\rho^{\sy}_N (\theta, T)$ in this sense, but has $N$ peaks adding an
oscillatory modulation to the function.  In some sense
$\rho^{\tr}_N(\theta,t)$ is intermediate between
$\rho^{\as}_N(\theta,t)$ and $\rho^{\sy}_N(\theta,t)$, since it can be
obtained from the expectation value of the ratio of values of the
characteristic polynomial evaluated at two different values of its
argument. The oscillatory behavior is in this sense a remnant of the
$\delta$-function structure of $\rho^{\as}_N(\theta,t)$. For this
reason we expect the peaks of $\rho^{\tr}_N(\theta,t)$ to occur at
locations close to the matching $\theta_j (\tau)$.  This expectation
will be confirmed numerically in section~\ref{sec:compare}.

Unlike for $\ell=\as$, $\sy$, explicit formulas for
$\rho^{\tr}_N(\theta,t)$ were unavailable so far. New formulas that
apply in this case will be derived in relative detail in
section~\ref{sec:true}.  We shall see that again a symmetry under
$N\to -N$ exists.

\section{\boldmath Motion of the zeros $z_j(\tau)$ as a function of
  $\tau$}
\label{sec:asym}

In this section we only consider $\rho^\as_N(\theta,\tau)$ and study
the zeros $z_j(\tau)=\exp(i\theta_j(\tau))$ of the average
characteristic polynomial for small, large, and near the critical
$\tau$.

\subsection[$\theta_j(\tau)$ for small $\tau$]{\boldmath
  $\theta_j(\tau)$ for small $\tau$}

\subsubsection{Approximate ``equations of motion''}

From equation~(\ref{eqmm}) we obtain
\begin{equation}
  \dot \theta_j =\frac{1}{N}\sum_{k\ne j} \sum_{n\in \Z}
  \frac{1}{\theta_j-\theta_k+2\pi n}\,.
  \label{eqm}
\end{equation}
Rescaling
\begin{equation}
  \theta_j=\frac{\eta_j}{\sqrt{N}}
\end{equation}
yields
\begin{equation}
  \dot \eta_j =\sum_{k\ne j} \sum_{n\in \Z}
  \frac{1}{\eta_j-\eta_k+2\pi n\sqrt{N}}\,. 
\end{equation}
The initial condition $\theta_j(\tau=0)=0$ indicates that one can
neglect to leading order in $\tau$ the terms with $n\ne 0$,
\begin{equation}
  \dot \eta_j \approx \sum_{k\ne j}  \frac{1}{\eta_j-\eta_k}\,.
\end{equation}
In this approximation periodicity under $\theta_j \to \theta_j +2\pi$
is lost, making the approximation unreliable when periodicity becomes
relevant.  This weak-coupling feature is a recurrent theme in models
that have compact variables and become disordered at strong couplings.

\subsubsection{Solution of the approximate equations}

Assigning dimension 1 to $\tau$ we see that $\eta$ has dimension 1/2.
We define
\begin{equation}
  \eta_j=\hat\eta_j \sqrt{2\tau}\,,
\end{equation}
making the $\hat\eta_j$ variables dimensionless and therefore
independent of $\tau$. They are determined by the equations
\begin{equation}
  \hat\eta_j =\sum_{k\ne j} \frac{1}{\hat\eta_j -\hat\eta_k}\,.
\end{equation}
The solution of these equations is well known, see, e.g.,
\cite[App.~A.6]{Mehta:1991}.  The $\hat\eta_j$ are the distinct zeros
of the Hermite polynomial $H_N(x)$,
\begin{equation}
  H_N(\hat\eta_j )=0\,,\qquad j=0,\ldots,N-1\,.
\end{equation}

\subsubsection{Relation to harmonic oscillator}

In the theory of orthogonal polynomials, the zeros of orthogonal
polynomials are shown to be the eigenvalues of the Jacobi matrix,
which is the appropriately truncated matrix of recurrence coefficients
\cite[Secs.~2.4 and 2.11]{Wilf:1978}. Introduce the matrix $a_N$, an
$N$-truncated version of the infinite dimensional annihilation
operator $a$ normalized by
\begin{equation}
  [a,a^\dagger]=1\,.
\end{equation}
The truncation is to the space spanned by the harmonic oscillator
states ${(a^\dagger)}^j |0\rangle$ with $j=0,\ldots,N-1$,
\begin{equation}
  a_N =\begin{pmatrix}0&\sqrt{1}&0&0&\cdots&0&0\cr
    0&0&\sqrt{2}&0&\cdots&0&0\cr
    \vdots&\vdots&\vdots&\vdots&\vdots&\vdots&0\cr
    0&0&0&0&\cdots&0&\sqrt{N-1}\cr
    0&0&0&0&\cdots&0&0\end{pmatrix}\,.
\end{equation}
The $a_N$ satisfy
\begin{equation}
  [a_N,a_N^\dagger ] = \1_N - N P_{N-1}\,,
\end{equation}
where $P_n=|n\rangle\langle n|$. 

Using the recurrence relations of the Hermite polynomials, the Jacobi
matrix is found to be $(a_N+a_N^\dagger)/\sqrt2$.  Thus, to leading
order in $\tau$ the zeros of $\langle\det(z-W)\rangle$ are the same as
the zeros of
\begin{equation}
  \det\left[z-e^{i\sqrt{\frac{\tau}{N}} (a_N + a_N^\dagger )}\right ]\,.
\end{equation}

\subsubsection{Largest zeros}
\label{sec:largest}

Of particular interest are the largest zeros in absolute magnitude.
They come in a pair of opposite signs.  Using a known formula for
large $N$ \cite[Eq.~(6.32.5)]{Szego:1991}, we have
\begin{equation}
  \hat\eta_M =\sqrt{2N}-\frac{1.856}{(2N)^{1/6}}+\ldots
\end{equation}
giving the largest $\theta_j$ as
\begin{equation}
  \theta_M(\tau)=2\sqrt{\tau} \left (1 -
    \frac{1.856}{(2N)^{2/3}} + \ldots \right )\,,\qquad
  M\equiv N-1\,.
\end{equation}
We now are in a position to estimate when $\tau$ cannot be considered
to be small anymore and the approximation first breaks down.

In~(\ref{eqm}) set $j=M$ and choose $k$ so that
$\theta_k=-\theta_M$. We see that by keeping only the $n=0$ term in
the sum we neglected, for example, the following potentially large
term,
\begin{equation}
  \frac{1}{2\theta_M-2\pi}\,.
\end{equation}
This is the point where ignoring periodicity becomes unacceptable.  At
$N\gg 1$ our small-$\tau$ approximation breaks down for
\begin{equation}
  2\sqrt{\tau}\left(1-\frac{1.856}{(2N)^{2/3}}\right)\approx\pi\,.
\end{equation}
In conclusion, the small-$\tau$ approximation holds for
\begin{equation}
  \sqrt{\tau} \ll \frac{\pi}{2}
\end{equation}
if $N\gg1$, but extends further if $N$ is not too large.  Since we
know that at infinite $N$ there is a transition at $\tau=4$, we see
that the small-$\tau$ approximation cannot take us all the way to the
critical point for $N\gg1$.

The most important conclusion is that the expansion in scale for small
loops yields a spectrum restricted to a finite arc centered at zero
angle and that the boundaries of the arc approach their infinite-$N$
limits by a leading term of order $N^{-2/3}$. This exponent is a
well-known property of the Gaussian ensemble of Hermitian matrices,
and is connected to universal functions constructed out of the Airy
function. The Airy function is in turn familiar from WKB wave
functions at linear turning points.  The power of 3 that appears in
the exponent of its integral representation is related to the
denominator 3 in the power of $N$ we just saw.

As the scale of the loop grows, the boundaries of the arc expand,
until they meet each other at $\theta=\pm\pi$, at which point the
small-scale expansion breaks down and the exponent changes.

\subsection[$\theta_j(\tau)$ for large $\tau$]{\boldmath
  $\theta_j(\tau)$ for large $\tau$}

\subsubsection[The eigenvalues at $\tau=\infty$]{\boldmath The
  eigenvalues at $\tau=\infty$}

The eigenvalues expand away from zero until they stop at
$\tau=\infty$, at which point they are equally spaced and contained in
the interval $(-\pi,\pi)$. Throughout the expansion they maintain the
sum rule
\begin{equation}
  \sum_{j=0}^{N-1} \theta_j (\tau )=0\,.
  \label{sumrule}
\end{equation}
This determines their asymptotic limits,
\begin{equation}
  \theta_j(\tau=\infty) = \frac{2\pi}{N}\left(j-\frac{N-1}{2}\right )
  \equiv \Theta_j\,,\qquad j=0,\ldots,N-1\,.
\end{equation}
We now prove that the above configuration is an equilibrium point in
the sense that the $\tau$-derivatives of the $\theta_j(\tau )$ vanish
for $\theta_j =\Theta_j$, $j=0,\ldots,N-1$. Since
\begin{equation}
  \dot \theta_j =-\frac{i}{2N}\sum_{k\ne j} \frac{1+e^{i(\theta_j-\theta_k)}}{1-
    e^{i(\theta_j-\theta_k)}}
\end{equation}
we need to show that for each $j=0,\ldots,N-1$
\begin{equation}
  \sum_{k\ne j} \frac{1+e^{i(\Theta_j-\Theta_k)}}{1-
    e^{i(\Theta_j-\Theta_k)}}=0\,.
  \label{claima}
\end{equation}
Let us denote by $q$ the $N$-roots of unity. A sum over $q$ runs over
these $N$ complex numbers. We need to show that
\begin{equation}
  A=\sum_{q\ne 1} \frac{1+q}{1-q}=0\,.
\end{equation}
This would then imply~(\ref{claima}). The above equation already
implies that the LHS of~(\ref{claima}) is independent of $j$.
Dividing by $q$ the numerator and denominator of the summand and
noticing that the restriction $q\ne 1$ is identical to the restriction
$1/q\ne 1$ for the $N$-roots of unity $q$, we get $A=-A=0$.

However, we shall soon need to evaluate other sums over $q$, and for
these a more general procedure is needed.  This procedure, when
applied to the present trivial case, goes as follows.  Start from
\begin{equation}
  A=\lim_{x\to 1^-} \left [ \sum_q \left ( \frac{1+xq}{1-xq} \right )
    -\frac{1+x}{1-x}\right ]\,.
\end{equation}
Next,
\begin{align}
  A&=\lim_{x\to 1^-} \biggl [ -N +2\sum_{n\ge 0} x^n \sum_q q^n
  -\frac{1+x}{1-x}\biggr ]=\lim_{x\to 1^-} \biggl [ -N +2N\sum_{k\ge
    0} x^{kN}-\frac{1+x}{1-x}\biggr ]\cr&= \lim_{x\to 1^-} \biggl [ -N
  +\frac{2N}{1-x^N}-\frac{1+x}{1-x}\biggr]
  =\lim_{\epsilon\to0^+}\biggl[-N+\left(\frac2\epsilon+N-1\right)
  -\frac{2-\epsilon}\epsilon\biggr] =0\,.
\end{align}
This again proves~(\ref{claima}).  Above, we observed that $\sum_q
q^n$ will be zero if $n$ is not a multiple of $N$, and $N$
otherwise. We need $x<1$ to perform the expansion in a geometric
series, but at the end we can take $x\to 1$.  Similar techniques work
for all other sums over $q$ we shall need.

\subsubsection[Linearization of the large-$\tau$ equation]{\boldmath
  Linearization of the large-$\tau$ equation}

We now expand around the infinite-$\tau$ solution, to see how it is
approached. From the exact formula for $\langle \det(z-W)\rangle$
in~\cite{Neuberger:2008mk}, we expect the approach to be exponentially
rapid, with decay constants given by the Casimirs of the antisymmetric
representations labeled by $l$, where $l=1,\ldots,N-1$. This is $N-1$
nonzero values, not $N$. The missing value corresponds to a uniform
$\tau$-independent shift in all $\theta_j(\tau )$, which is a symmetry
of the differential equation. This symmetry would produce a zero mode
in the linearized equation, but the mode is eliminated by the sum
rule~(\ref{sumrule}), which depends also on the initial condition.

To linearize we set
\begin{equation}
  \theta_j(\tau)=\Theta_j +\delta \theta_j (\tau)
\end{equation}
and expand the equation of motion to linear order in $\delta
\theta_j$.  Unlike the initial condition, the set $\{\Theta_j\}$
provides a nondegenerate configuration around which it is
straightforward to expand. We find
\begin{equation}
  \delta\dot \theta_j =-\frac{1}{4 N} \sum_{k=0}^{N-1} A_{jk}\delta
  \theta_k 
\end{equation}
with the matrix $A$ given by
\begin{equation}
  A_{jk}=
  \begin{cases}
     -\frac{1}{\sin^2\frac{\Theta_j-\Theta_k}{2}} &
     \text{for }k\ne j\,,\\
     \sum\limits_{k\ne j}\frac{1}{\sin^2\frac{\Theta_j-\Theta_k}{2}}&
     \text{for }k=j\,.  
   \end{cases}
\end{equation}
We need the eigenvalues and eigenvectors of this matrix.  Note first
that $A_{jj}$ does not depend on $j$ and is given by
\begin{equation}
  A_{jj}=4\sum_{q\ne 1} \frac{1}{(1-q)(1-q^{-1})}\,.
\end{equation}
The sum over $q$ can be performed as before leading to
\begin{equation}
  A_{jj}=\frac{N^2-1}{3}\,.
\end{equation}
Hence, the matrix $A$ has entries $A_{ij}$ which only depend on
$(i-j)\bmod N$.  Therefore, $A$ has $N$ eigenvectors $\phi^{(l)}$ with
components $\phi^{(l)}_k$, $k,l=0,\ldots,N-1$, given by
\begin{equation}
  \phi^{(l)}_k =\frac1{\sqrt N} e^{-i\frac{\pi l (N-1)}{N}}
  e^{i\frac{2\pi l}{N} k}\,.
\end{equation}
The phases have been chosen for later convenience.  To evaluate the
action of $A$ on an eigenvector $\phi^{(l)}$, we need to perform sums
of the type
\begin{equation}
  \xi^{(l)}=-4\sum_{q\ne 1} \frac{q^l}{(1-q)(1-q^{-1})}\,.
\end{equation}
The sum over $q$ is performed as before, and one gets
\begin{equation}
  \xi^{(l)}+\frac{N^2-1}{3}=2l(N-l)\,.
\end{equation}
The RHS is the eigenvalue of $A$ corresponding to the $l$-th
eigenvector $\phi^{(l)}$. $l=0$ corresponds to the zero mode which
does not contribute to the $\delta \theta_j$, so we are left with
$N-1$ contributing modes, labeled by $l=1,\ldots,N-1$.  As expected,
the eigenvalues of $A$ come out proportional to the quadratic Casimirs
in the $l$-fold antisymmetric representation, given by
\cite{Perelomov:1965ab}
\begin{align}
  C_2(l)=\frac{N+1}Nl(N-l)\,.
\end{align}
The equations of motion~(\ref{eqmm}) have the values of the Casimirs
encoded in them.

Thus we have found that
\begin{equation}
  \delta \theta_k (\tau )=\sum_{l=1}^{N-1} C_l \phi^{(l)}_k
  e^{-\frac{\tau}{2N} l(N-l)}\,. 
\end{equation}
It remains to determine the coefficients $C_l$.  Since the leading
asymptotic terms at large $\tau$ correspond to $l=1$ and $l=N-1$, we
only need $C_1$.

\subsubsection{Constraints on the coefficients}

The coefficients $C_l$ are restricted by two quite trivial exact
general properties, which imply for the $\delta \theta_k (\tau )$ that
\begin{align}
  \delta \theta_k (\tau )&= \delta \theta_k^\ast (\tau)\,,\qquad
  \delta \theta_k (\tau )= -\delta \theta_{N-k-1} (\tau )\,.
\end{align}
These constraints lead to
\begin{equation}
  \delta \theta_k (\tau)= \sum_{l=1}^{N-1} \rho_l 
  \sin\left [ \frac{2\pi l}{N} (k+1/2) \right ]
  e^{-\frac{\tau}{2N} l(N-l)}
  \label{eq:con}
\end{equation}
with real $\rho_l$ and
\begin{equation}
  \rho_l=\rho_{N-l}\,.
\end{equation}
Every term in the sum representing $\delta \theta_k (\tau )$ is
invariant under $l\to N-l$.

\subsubsection{Leading asymptotic behavior}

Note first that we have
\begin{align}
  \langle \Tr W \rangle = \sum_{k=0}^{N-1}e^{i\theta_k}\,,
\end{align}
which is the term proportional to $z^{N-1}$ in the expansion of
\eqref{eq:det} in $z$.

For the leading asymptotic behavior of the $\theta_k(\tau)$ we only
need $\rho_1$. We can obtain $\rho_1$ from the exact result
\begin{equation}
  \frac{1}{N} \langle \Tr W \rangle = e^{-\frac{\tau}{2N} (N-1)}\,.
\end{equation}
Actually, we only need this result at leading order as
$\tau\to\infty$.  To linear order in $\delta \theta_k$, and keeping
only the terms with $l=0$ and $l=N-1$ in \eqref{eq:con}, we have
\begin{equation}
  \label{eq:lead}
  \frac{1}{N} \langle \Tr W \rangle =-\frac{2 i}{N} \rho_1 \sum_{k=0}^{N-1}
  e^{\frac{2\pi i}{N} (k+1/2) } \sin\left [ \frac{2\pi}{N}
    (k+1/2)\right ]  e^{-\frac{\tau}{2N} (N-1)}\,.
\end{equation}
Performing the trivial sum over $k$ we get
\begin{equation}
  \rho_1=1\,.
\end{equation}
Hence, as $\tau\to\infty$ 
\begin{equation}
  \delta \theta_k (\tau) \sim 2\sin\left [ \frac{2\pi}{N}
    (k+1/2)\right ]  e^{-\frac{\tau}{2N} (N-1)}
\end{equation}
or, more completely,
\begin{equation}
  \theta_k (\tau ) \sim \frac{\pi}{N} (2k+1 -N ) + 
  2\sin\left [ \frac{2\pi}{N}
    (k+1/2)\right ]  e^{-\frac{\tau}{2N} (N-1)}\,.
\end{equation}
Equivalently, we can write
\begin{equation}
  \theta_k (\tau ) \sim \Theta_k - 2   e^{-\frac{\tau}{2N} (N-1)} 
  \sin(\Theta_k )\,.
\end{equation}
For $\Theta_k$ negative the correction is positive and for $\Theta_k$
positive the correction is negative.  This shows that, as expected,
for increasing $\tau$ each eigenvalue is distancing itself from the
origin for all $k$.  The correction is largest for eigenvalues in the
middle of the upper and lower half of the circle -- the eigenvalues
here are the last to settle into their infinite-$\tau$ destinations.

\subsection[Extremal $\theta_j (\tau)$ for $\tau\sim 4$ and large
$N$]{\boldmath Extremal $\theta_j (\tau)$ for $\tau\sim 4$ and large
  $N$}

\subsubsection{Universal zeros}

In terms of the variable $y$ from~\cite{Neuberger:2008mk}, the zeros
corresponding to the angles $\theta_j (\tau)\bmod2\pi$ are given by
\begin{equation}
  \hat{q}_N \left ( i(\theta_j(\tau)-\pi ),\tau\right )=0
\end{equation}
with
\begin{equation}
  \hat{q}_N (y,\tau)=\int_{-\infty}^{\infty} dx \,e^{-\frac{N}{2\tau}
    (y-x)^2} e^{N\log(2\cosh (x/2))}\,. 
\end{equation}
The universal form of $\hat q_N (y,\tau)$ for large $N$, $y\sim 0$,
and $\tau \sim 4$ is obtained by replacing the $\log(2\cosh(x/2))$
above by its expansion truncated at order $x^4$,
\begin{equation}
  \log\left (\cosh\frac{x}{2}\right ) =\frac{x^2}{8} -\frac{x^4}{192}\,.
\end{equation}
At $\tau=4$, we have
\begin{equation}
  \hat{q}_N (y,4)=\int_{-\infty}^{\infty} dx \, e^{-\frac{N}{8}(x-y)^2
  } e^{N\log(2\cosh (x/2))}\,. 
\end{equation}
The ``universal zeros'' $y_\ast^j$ are defined by 
\begin{equation}
  \int_{-\infty}^{\infty} dx \, e^{-\frac{N}{192} (x^4 - 48 xy_\ast^j
    )}=0\,. 
\end{equation}

\subsubsection{Universal numerical values}

Universal zeros have been investigated in~\cite{Senouf:1996ab}.
Define
\begin{align}
  \frac{Nx^4}{192}=\mu u^4\,,\qquad \frac{Nxy_*^j}{4}=4i\mu u\,.
\end{align}
Then
\begin{align}
  y_*^k = \pm i\frac{4\sqrt2}{3}\left(\frac{3\mu_k}N\right)^{3/4}\,,
\end{align}
where the $\mu_k$, $k=1,2,\ldots$ are the zeros of
\begin{equation}
  F(\mu)=\int du \, e^{\mu(4iu - u^4 )}\,.
\end{equation}
From Table 1 of~\cite{Senouf:1996ab}, we have $\mu_1=0.8221, \mu_2
=2.0227,\ldots$ Various other results concerning the $\mu_k$ can be
found in~\cite{Senouf:1996ab}.  For the extremal positive zero at
$\tau=4$ we need to look at $y_*^1$,
\begin{equation}
  y^1_* \approx i \frac{3.7}{N^{3/4}}\,.
\end{equation}
This gives, for large $N$, that the zero $z_j(\tau_c)$ that is closest
to $-1$ with $\im z_j >0$ is
\begin{equation}
  z_M \approx e^{i(\pi -3.7 N^{-3/4})}\,.
  \label{zMcritical}
\end{equation}

\section{\boldmath Asymptotic expansion of $\rho^{\sy}_N (\theta, T)$}
\label{secSymmSaddle}

The aim of this section is to construct an asymptotic expansion of
$\rho_N^\sy(\theta,T)$ in powers of $1/N$.  To this end we perform a
saddle-point analysis of the integral in \eqref{psint}, from which
$\rho_N^\sy(\theta,T)$ can be obtained via \eqref{eq:phipm} and
\eqref{eq:rhosym}.  It is sufficient to study only $\psi_+^{(N)}(z,T)$
because $\psi_-^{(N)}(z,T)$ can be obtained from \eqref{eq:psi-}.

\subsection{Saddle-point analysis}
\label{sec:spa}

For $|z|=1$ the integrand of \eqref{psint} has singularities on the
real-$u$ axis.  We therefore set $z=e^{\epsilon+i\theta}$, where
$\epsilon >0$ ensures that $|z|>1$ but will later be taken to zero.
The integrand of \eqref{psint} can be written as $\exp(-Nf(u))$ with
\begin{align}
  f(u)=\frac{u^2}{2T}+\log\left(ze^{-i\frac u2}-e^{i\frac u2}\right)\,.
\end{align}
We now look for saddle points of the integrand in the complex-$u$
plane, which we label by $\bar u=iTU(\theta,T)$, where
$U(\theta,T)=U_r (\theta,T) + i U_i (\theta, T)$ is a complex-valued
function of $\theta$ and $T$.  The saddle-point equation turns out to
be
\begin{equation}
  e^{-TU(\theta,T)}\frac{U(\theta, T) + 1/2}{U(\theta, T) -1/2}
  =e^{\epsilon+i\theta}\,.
  \label{contour}
\end{equation}
For $\epsilon=0$, this is equation (5.49) in~\cite{Lohmayer:2008bd}
and is related to the inviscid complex Burgers equation via equation
(5.44) there. In the present notation, the latter equation has the
form
\begin{equation} 
  \frac{\partial U}{\partial T}+iU\frac{\partial
    U}{\partial \theta} =0\,.
\end{equation}
Taking the absolute value of \eqref{contour} leads to the equation
\begin{align}
  \label{eq:uiur}
  U_i^2 &= U_r\coth (T U_r+\epsilon)-U_r^2 -\frac{1}{4}\,.
\end{align}
For $\epsilon=0$, this equation has been investigated previously
in~\cite{Lohmayer:2008bd}.  However, here we keep $\epsilon>0$ for the
time being.  The singularities of the integrand of \eqref{psint} then
all have $U_r<0$.  Equation~\eqref{eq:uiur} describes one or more
curves in the complex-$U$ plane on which the saddle points have to lie
(for a given value of $\theta$, the saddles are isolated points on
these curves). %
\FIGURE[t]{
  \includegraphics[width=.3\textwidth]{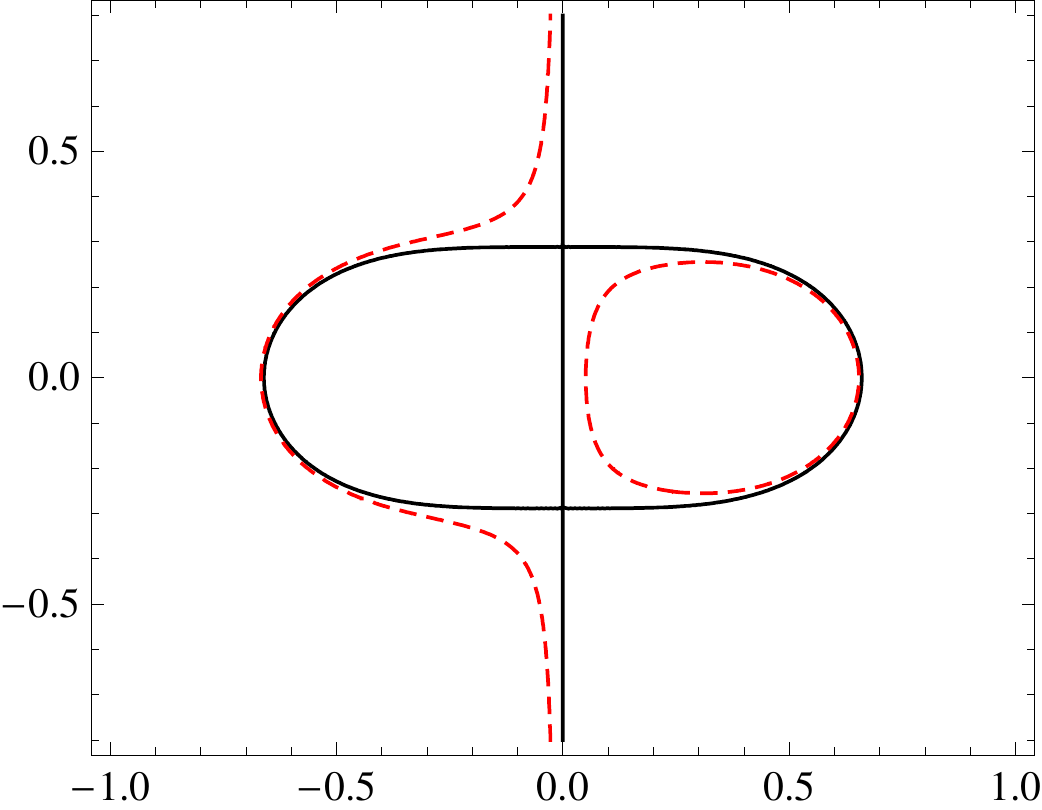}\hfill
  \includegraphics[width=.3\textwidth]{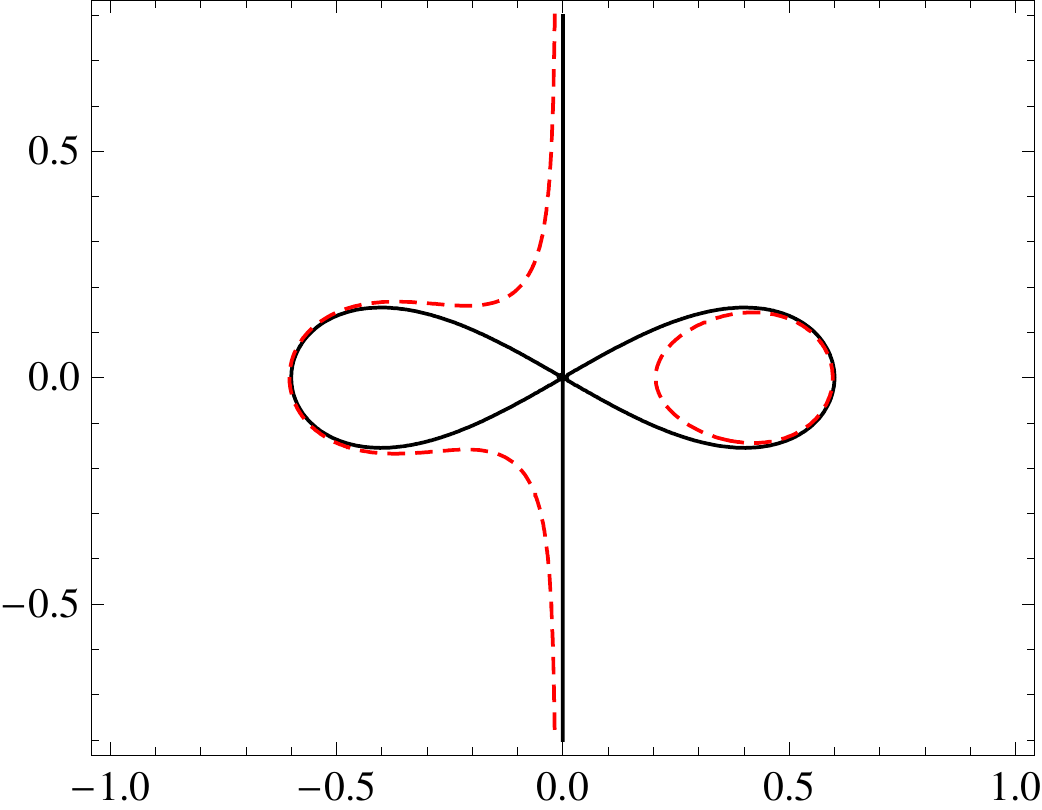}\hfill
  \includegraphics[width=.3\textwidth]{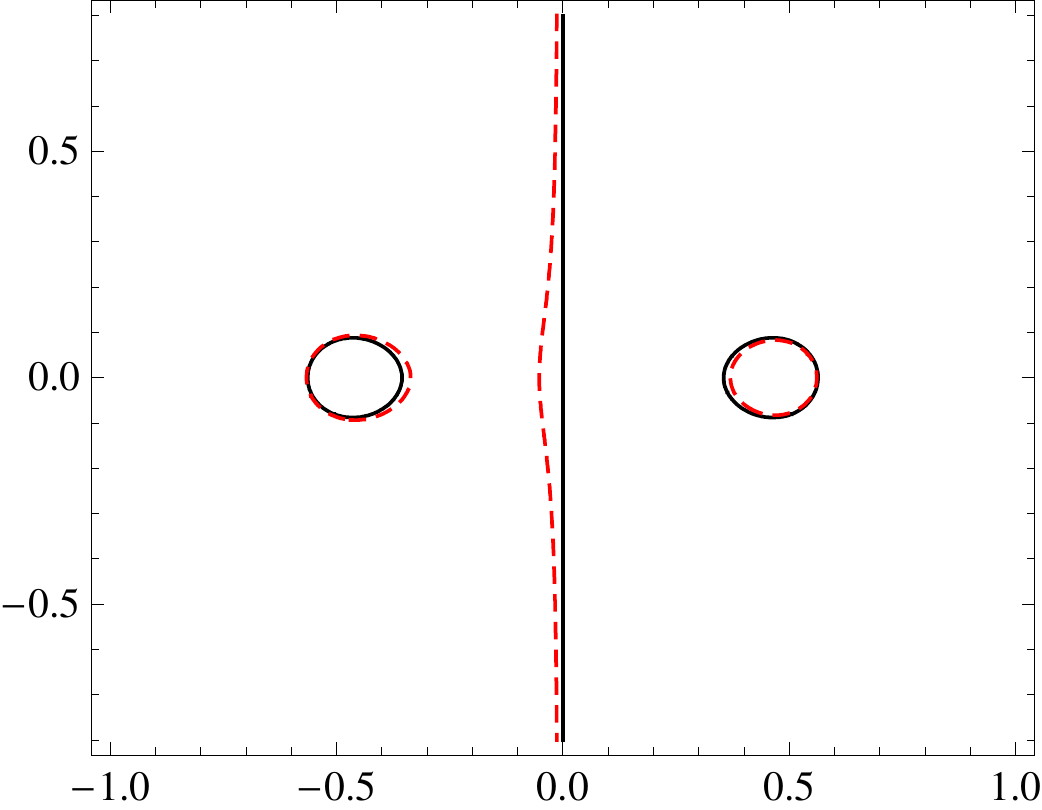}
  \caption{Examples of the contours in the complex-$U$ plane described
    by equation~\eqref{eq:uiur} for $T=3$ (left), $T=4$ (middle), and $T=5$
    (right).  The red dashed curves are for small $\epsilon>0$, while
    the solid black curves are for $\epsilon=0$.  For our saddle-point
    analysis we keep $\epsilon>0$.}
  \label{fig:curves}
}%
In figure~\ref{fig:curves} we show typical examples for these curves
for $T<4$, $T=4$, and $T>4$, where $\epsilon$ has been chosen
sufficiently close to zero.  (The closed contours always enclose the
points $U=1/2$ or $U=-1/2$.  For $T>4$ and larger $\epsilon$, the
closed contour in the left half-plane would be missing, but right now
we are not concerned with this since we are only interested in the
limit $\epsilon\to0^+$.)  Analyzing \eqref{contour} numerically we
find, for all values of $T$, that for a given value of $\theta$ there
is always one (and only one) saddle point on the closed contour in the
right half-plane, i.e., with $U_r>0$.  Note that we are showing the
complex-$U$ plane, in which the original integration contour
corresponds to the imaginary axis.  The integration contour can be
smoothly deformed to go through the (single) saddle point in the right
half-plane along a path of steepest descent.  No singularities are
crossed since they all have $U_r<0$.  There are also saddle points on
the contour(s) in the left half-plane (in fact, there are infinitely
many on the open contour), but these need not be considered.

Once the integration contour has been deformed to go through the
saddle point, we can safely take the limit $\epsilon\to0^+$.
Parametrizing the contour in the vicinity of the saddle point by
$u=\bar u+xe^{i\beta}$, where $x$ is the new integration variable
corresponding to the fluctuations around the saddle and $\beta$ is the
angle which the path of steepest descent makes with the real-$u$ axis,
$\psi^{(N)}_+ (e^{i\theta},T)$ is given, up to exponentially small
corrections in $N$, by
\begin{align}
  \psi^{(N)}_+ (e^{i\theta},T) &= \frac1{2^N} \sqrt{\frac{N}{2\pi
      T}}\, e^{\frac{NT}{8}-i\frac{N\theta}{2}+i\beta}
  \int_{-\infty}^\infty
  dx \, e^{-Ng(x)}\,,\\
  g(x)&=\frac1{2T}\bigl(xe^{i\beta}+iTU(\theta,T)\bigr)^2
  +\log\sinh\frac{i\theta-ixe^{i\beta}+TU(\theta,T)}{2}\,.
\end{align}
We can now expand $g(x)$ in $x$.  The linear order vanishes by
construction.  The second order gives a Gaussian integral over $x$,
resulting in
\begin{align}
  \psi^{(N)}_+ (e^{i\theta},T)&\approx 
  e^{\frac{NT}{8}+\frac{N T U^2 (\theta, T)}2}
  \frac{\left[e^{-i\theta}(1/4-U^2(\theta,T))\right]^{N/2}}
  {\sqrt{1-T(1/4-U^2(\theta,T))}} \,.
  \label{eq:spgauss}
\end{align}
Note that the factor $e^{-i\theta}$ cannot be pulled out of the term
in square brackets because periodicity in $\theta$ would be lost.

There is a potential complication.  In principle, $g''(0)$ and
therefore the denominator in \eqref{eq:spgauss} could be zero, which
would mean that the integral over $x$ cannot be performed in Gaussian
approximation.  For $T>4$, it is straightforward to show that $g''(0)$
is never zero.  For $T\le4$, one can use \eqref{contour} to show that
$g''(0)=0$ only for the saddle points corresponding to the two angles
$\theta=\pm\theta_c(T)$ at which $\rho_\infty(\theta,T)$ becomes zero
(see section~\ref{subsecRhoSymm}).  This means that for
$|\theta|=\theta_c(T)$ the asymptotic expansion in $1/N$ diverges, and
that it converges ever more slowly as $|\theta|\to\theta_c$ from
below.

Note that for $T<4$ and $\theta_c(T)\le|\theta|\le\pi$ the function
$\rho^\sy(\theta,T)$ is exponentially suppressed in $N$.  The study of
the large-$N$ asymptotic behavior in this region requires more work.

\subsection{Leading-order result}

Equation~\eqref{eq:spgauss} is the leading order in the $1/N$
expansion of $\psi^{(N)}_+ (e^{i\theta},T)$.  We now show that it
leads to $\rho_N^\sy(\theta,T)\to\rho_\infty(\theta,T)$ as
$N\to\infty$.  We first write \eqref{eq:spgauss} in the form
\begin{align}
  \frac1N\log\psi_+^{(N)}(e^{i\theta},T)=\frac T8-f(\bar u)+\mathcal
  O(1/N)\,.
\end{align}
Note that in this order we do not need the denominator in
\eqref{eq:spgauss}, which corresponds to $f''(\bar u)$ (or $g''(0)$).
Via \eqref{eq:phipm} and using $\bar u=iTU$ this leads to
\begin{align}
  \phi_+^{(N)}(z,T)=i\left(\frac12-\frac z{z-e^{-TU}}\right)+\mathcal
  O(1/N)=-iU+\mathcal O(1/N)\,,
\end{align}
where in the last step we have used the saddle-point equation
\eqref{contour}.  Equation \eqref{eq:rhosym} then gives
\begin{align}
  \lim_{N\to\infty}\rho_N^\sy(\theta,T)=2\re U(\theta,T)\,,
\end{align}
which equals $\rho_\infty(\theta,T)$ of DO
\cite{Durhuus:1980nb,Janik:2004tw} since $U(\theta,T)$ satisfies
\eqref{contour} (which leads to \eqref{SaddleLambda} below with
$\lambda=U-1/2$ and $v=1/z$).

\subsection[$1/N$ correction to $\rho_\infty(\theta,T)$]{\boldmath
  $1/N$ correction to $\rho_\infty(\theta,T)$}
\label{sec:spa1N}

Higher-order terms in the $1/N$ expansion of $\psi^{(N)}_+
(e^{i\theta},T)$ can be obtained in the standard way by considering
higher powers of $x$ in the expansion of $g(x)$, resulting in
integrals of the type $\int_{-\infty}^\infty
dx\,x^{2n}e^{-g''(0)x^2/2}$ with $n\in\mathbb{N}$.  However, if we are
only interested in the $1/N$ correction to $\rho_\infty(\theta,T)$ the
result \eqref{eq:spgauss} is already sufficient ($1/N$ corrections to
this result would give $1/N^2$ corrections to
$\rho_\infty(\theta,T)$).  Therefore we now write
\begin{align} 
  \frac1N\log\psi_+^{(N)}(e^{i\theta},T)=\frac T8-f(\bar u)
  -\frac1{2N}\log[Tf''(\bar u)]+\mathcal O(1/N^2)\,,
\end{align}
which leads to
\begin{align}
  \phi_+^{(N)}(z,T)=-iU\left(1+\frac 1N
    \frac{T(1/4-U^2)}{[1-T(1/4-U^2)]^2}\right) +\mathcal O(1/N^2)
\end{align}
and thus to
\begin{align}
  \label{eq:rhosym1N}
  \rho_N^\sy(\theta,T)=2\re \left[U \left(1+\frac 1N
      \frac{T(1/4-U^2)}{[1-T(1/4-U^2)]^2}\right)\right]+\mathcal
  O(1/N^2) \,.
\end{align}
Note that for $T\le4$ and $|\theta|\to\theta_c(T)$ (from below) the
denominator of the $1/N$ term approaches zero, which corresponds to
the complication discussed in section~\ref{sec:spa}.  Note also that
for $T\le4$ and $|\theta|>\theta_c$ the saddle point $U(\theta,T)$ is
purely imaginary so that both the leading order and the $1/N$ term are
zero.  This confirms that the above saddle-point analysis is not the
right tool to compute finite-$N$ effects in this region.

In figure~\ref{fig:1N} we show examples for the $1/N$ corrections to
$\rho_\infty(\theta,T)$ for $N=10$ and $T=2$ and $5$. %
\FIGURE[t]{
  \includegraphics[width=0.4\textwidth]{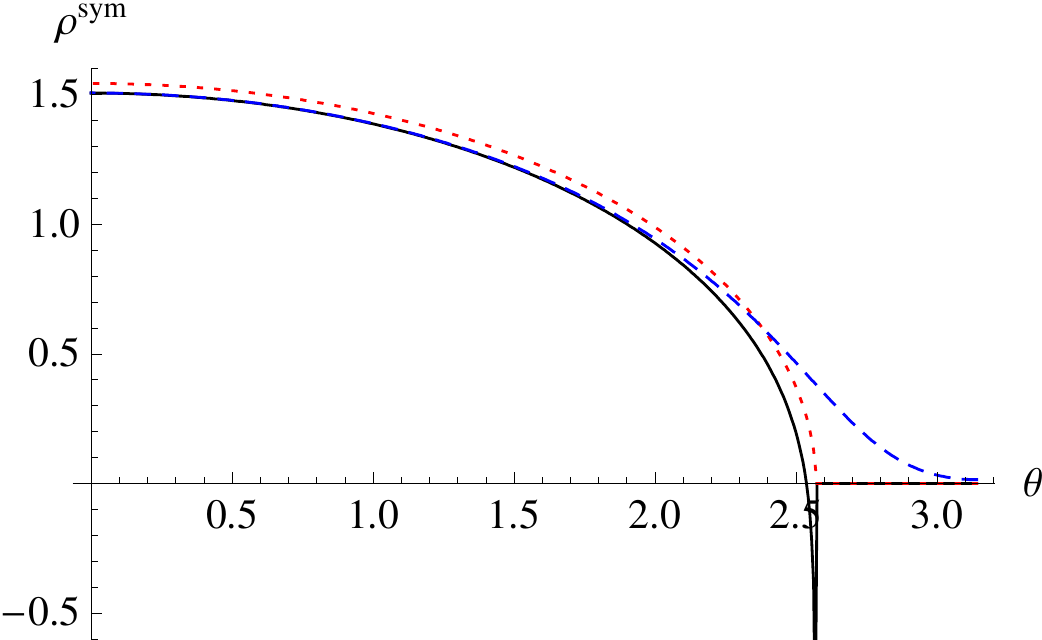}\hfill
  \includegraphics[width=0.4\textwidth]{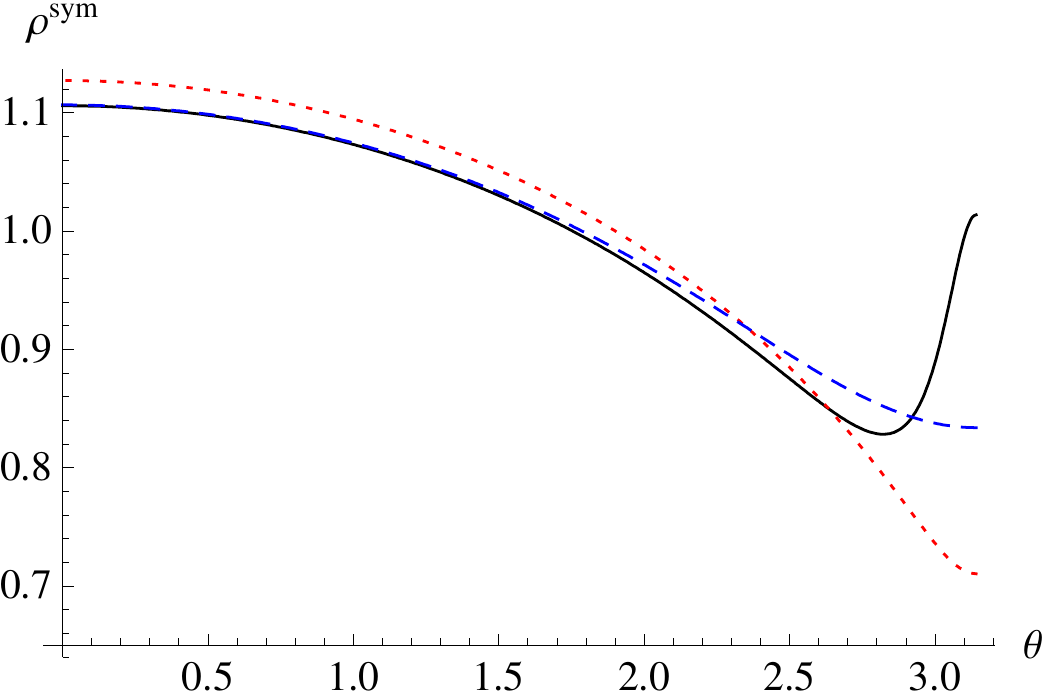}
  \caption{Examples for the $1/N$ corrections to
    $\rho_\infty(\theta,T)$ for $N=10$, $T=2$ (left), and $T=5$
    (right).  Shown are the exact result for $\rho_N^\sy(\theta,T)$
    (blue dashed curve), the infinite-$N$ result
    $\rho_\infty(\theta,T)$ (red dotted curve), and the asymptotic
    expansion of $\rho_N^\sy(\theta,T)$ up to order $\mathcal O(1/N)$
    from \eqref{eq:rhosym1N} (black solid curve).  We observe that the
    asymptotic expansion converges rapidly for small $|\theta|$ and
    more slowly for larger $|\theta|$.}
  \label{fig:1N}
}

\section{\boldmath The true eigenvalue density at finite $N$}
\label{sec:true}

We now proceed to derive exact formulas for the eigenvalue density
$\rho_N^\tr(\theta,t)$. 

\subsection{Character expansion}

To compute \eqref{eq:tr} we consider the ratio of determinants
\begin{equation}
  R(u,v,W) \equiv \frac{\det(1+uW)}{\det(1-vW)}
\end{equation}
with $|v|<1$ and expand it in $\SU(N)$ characters using \cite{Neuberger:2008ti}
\begin{align}
  \label{eq:exp}
  \det(1+uW)=\sum_{p=0}^N u^p\chi_p^A(W)\,,\qquad
  \det(1-vW)=\sum_{q=0}^\infty v^q\chi_q^S(W)\,,
\end{align}
where $\chi_p^A(W)$ ($\chi_q^S(W)$) denotes the character of $W$ in a
totally antisymmetric (symmetric) representation whose Young diagram
consists of a single column (row) with $p$ ($q$) boxes.  The trivial
representation corresponds to $p=0$ ($q=0$), and for $\SU(N)$ the
antisymmetric representation with $p=N$ boxes is equivalent to the
trivial one because of $\chi_N^A(W)=\det W=1$.  This yields
\begin{equation}
R(u,v,W)=\sum_{p=0}^N\sum_{q=0}^\infty u^p v^q \chi_p^A (W) \chi_q^S (W)\,.
\end{equation}
The task now is to decompose the tensor product $p^A\otimes q^S$ into
irreducible representations.  In general, $p^A\otimes q^S$ consists of
tensors with $p+q$ indices, where the first $p$ are antisymmetrized
and the last $q$ are symmetrized.  To decompose into irreducible
representations we take one index from the first $p$ and one from the
last $q$ and either symmetrize or antisymmetrize this pair.  There are
no more symmetrization operations we can perform. Thus,
\mbox{$p^A\otimes q^S$} decomposes into two irreducible
representations, except in boundary cases when it is already
irreducible. The boundary cases are at $q=0$ or $p=0$ or $p=N$. Away
from the boundary cases $p^A\otimes q^S$ decomposes into two
irreducible representations identified by Young diagrams with the top
row consisting of $h$ boxes and a left column of $v$ boxes and nothing
else:
\begin{equation}
  \label{eq:hook}
  \young(12\hfil\hfil h,2,\hfil,\hfil,v)
\end{equation}
One either has $h=q$ and $v=p+1$ or $h=q+1$ and $v=p$.  (Do not
confuse the $v$ here with the argument of $R$.)  The first case
corresponds to an antisymmetrized pair and the second to a symmetrized
pair.  For later convenience we shall label the ``hook'' diagram in
\eqref{eq:hook} by $(v-1,h-1)$, with the understanding that $v=0$ or
$h=0$ gives the trivial representation.  We thus have
\begin{align}
  p^A\otimes q^S=(p,q-1)\oplus(p-1,q)
\end{align}
and for the boundary cases
\begin{align}
  p^A\otimes0=(p-1,0)\,,\quad 0\otimes q^S = (0,q-1)\,,\quad N^A
  \otimes q^S=(N-1,q)=(0,q-1)\,.
\end{align}
Taking into account these boundary cases and suppressing the $\SU(N)$
matrix argument $W$, we obtain 
\begin{equation}
  R(u,v)=1+\sum_{p=0}^{N-1}\sum_{q=1}^\infty u^p v^q \chi_{(p,q-1)} +
  \sum_{p=1}^{N}\sum_{q=0}^\infty u^p v^q \chi_{(p-1,q)}\,.
\end{equation}
The case $p=0$, $q=0$ is excluded from the sums. Every other boundary case
appears in exactly one of the two sums above. Every nontrivial pair
has one of the two irreducible representations in exactly one of the
sums.  Now change summation indices $q\to q+1$ in the first sum and
$p\to p+1$ in the second to obtain
\begin{equation}
  R(u,v)=1+(u+v)\sum_{p=0}^{N-1}\sum_{q=0}^\infty u^p v^q
  \chi_{(p,q)}\,.
  \label{eq:dsum}
\end{equation}
This makes it explicit that $R=1$ at $u=-v$.

A consequence is the character expansion of $\Tr W^k$ for all $k$.
Since
\begin{equation}
  R(-v+\epsilon, v) = 1-\frac{N\epsilon}{v} +\frac{\epsilon}{v} \Tr
  \frac{1}{1-vW} +{\cal O}(\epsilon^2)\,, 
\end{equation}
we have 
\begin{equation}
  \Tr\frac{1}{1-vW} = N+v\sum_{(p,q)}(-1)^p{v^{p+q}}{\chi_{(p,q)} (W)}\,,
\end{equation}
where the limits on the double sum are given in \eqref{eq:dsum}.
Hence, taking $k>0$,
\begin{equation}
  \Tr W^k= \sum_{{(p,q)}\atop{p+q=k-1}} (-1)^p \chi_{(p,q)} (W)\,.
\end{equation}
Obviously, $\Tr 1=N$ and $\Tr W^{-k}= (\Tr W^k)^\ast$. 

\subsection{Performing the average}

The average over $W$ with weight \eqref{eq:weight} produces, using
character orthogonality,
\begin{equation}
  \langle \chi_{(p,q)} (W) \rangle = d(p,q) e^{-\frac{t}{2N} C(p,q)}\,,
\end{equation}
where $C(p,q)$ is the value of the quadratic Casimir operator in
$(p,q)$, given by \cite{Perelomov:1965ab}
\begin{equation}
  C(p,q)=(p+q+1)\left(N-\frac{p+q+1}{N}+q-p\right)
\end{equation}
and the dimension of the irreducible representation labeled by $(p,q)$
is
\begin{equation}
  \label{eq:dim}
  d(p,q)
  =d^A (p) d^S (q) \frac{(N-p)(N+q)}{N} \frac{1}{p+q+1}
\end{equation}
with
\begin{equation}
 d^A (p) = {N \choose p}\,,\qquad d^S(q) = {N+q-1\choose q}\,.
\end{equation}

\subsection{Basic combinatorial identities}

The expansions of one determinant or one inverse determinant factor
(i.e., setting $W=1$ and $u=\xi$, $v=0$ or $u=0$, $v=\eta$ in
\eqref{eq:dsum}) provide the identities 
\begin{subequations}\label{identity}
  \begin{align}
    \Sigma^A(\xi)&\equiv\sum_{p=0}^{N-1} \xi^p d^A(p) (N-p) =
    N(1+\xi)^{N-1}\,,\\ 
    \Sigma^S(\eta)&\equiv\sum_{q=0}^\infty \eta^q d^S (q) (N+q) =
    \frac{N}{(1-\eta)^{N+1}}\,,
  \end{align}
\end{subequations}
with $|\eta|<1$.
These will be needed to carry out the summations over $p$ and $q$
later. 

\subsection[Factorizing the sums over $p$ and $q$ for the average
resolvent at zero area]{\boldmath Factorizing the sums over $p$ and
  $q$ for the average resolvent at zero area}

Set $u=-v+\epsilon$. Up to corrections of order $\epsilon^2$ we have
\begin{equation}
  R(-v+\epsilon, v ,W)= 1 +\epsilon \sum_{p=0}^{N-1}\sum_{q=0}^\infty
  (-1)^p v^{p+q} \chi_{(p,q)} (W)=1-\epsilon
  \Tr\frac{1}{v-W^\dagger}\,. 
\end{equation}
This leads to
\begin{equation}
  \label{barRsum}
  \bar R(v)\equiv \braket{ \Tr\frac{1}{v-W^\dagger } } = -
  \sum_{p=0}^{N-1}\sum_{q=0}^\infty (-1)^p v^{p+q} e^{-\frac{t}{2N}
    C(p,q)} d(p,q)\,, 
\end{equation}
where $t=\lambda{\cal A}$. Note that the sum can be extended to $p=N$
because of the factor $N-p$ in $d(p,q)$.  Using \eqref{eq:tr} and
\eqref{eq:true}, we obtain
\begin{align}
  \label{eq:rhotrue}
  \rho^\tr(\theta,t)=1-\frac2N\lim_{\epsilon\to0^+}\re[v\bar R(v)]\,,
  \qquad v=e^{-\epsilon+i\theta}\,.
\end{align}
Note that there is no need for the limiting procedure $\epsilon\to0^+$
in \eqref{eq:rhotrue} if we are using the double sum in
\eqref{barRsum} for $\bar R(v)$, which is well-defined for $|v|=1$.

We now introduce an integral to get rid of the denominator in
\eqref{eq:dim} and obtain 
\begin{align}
  \bar R (v) &= -\int_0^1 \frac{d\rho}{N}
  \sum_{p=0}^{N}\sum_{q=0}^\infty \left [ (-1)^p v^p \rho^p d^A (p)
    (N-p)\right ] \left [ v^q \rho^q d^S (q) (N+q) \right ]\cr
  &\quad\times e^{-\frac{t}{2N} (p+q+1) (N - \frac{p+q+1}{N} + q -p )
  }\,.
\end{align}
This achieves factorization of the sums over $p$ and $q$ at $t=0$. The
sum in each factor can be performed using~(\ref{identity}).

\subsection{Integral representation at any area}

The $t$-dependent weight factor is the exponent of a bilinear form in
$p$ and $q$.  By a Hubbard-Stratonovich transformation the dependence
of the exponent on $p$ and $q$ can be made linear, and then the sums
over $p$ and $q$ are factorized for every $t$ and can again be done
exactly using (\ref{identity}).

Define the complex symmetric matrix $B_N$ by
\begin{equation}
  B_N =\begin{pmatrix} 1+\frac{1}{N} & \frac{i}{N} \cr \frac{i}{N} &
    1-\frac{1}{N}\end{pmatrix}\,. 
\end{equation}
$B_N$ has only one eigenvalue (equal to one) and is nondiagonalizable.
We have $\det B_N =1$ and
\begin{equation}
  B_N^{-1} = \begin{pmatrix} 1-\frac{1}{N} & -\frac{i}{N} \cr
    -\frac{i}{N} & 1+\frac{1}{N}\end{pmatrix}\,. 
\end{equation}
The quadratic Casimir form can be written with the help of $B_N$:
\begin{equation}
  C(p,q) = \begin{pmatrix}ip \\ q\end{pmatrix}^T B_N \begin{pmatrix}ip
    \cr q\end{pmatrix} +N\left(1-\frac{1}{N^2}\right) 
  +N\left(1+\frac{1}{N} - \frac{2}{N^2} \right) q 
  +N\left(1-\frac{1}{N} - \frac{2}{N^2} \right) p\,.
\end{equation}
Hence
\begin{align}
  e^{-\frac{t}{2N} C(p,q)} &=\frac{N}{t}\; e^{-\frac{t}{2}
    \left(1-\frac{1}{N^2}\right)}\;\int_{-\infty}^{\infty}
  \int_{-\infty}^{\infty} \frac{dx dy}{2\pi} \; \exp\left
    [-\frac{N}{2t} \begin{pmatrix}x&y\end{pmatrix}B_N^{-1} \begin{pmatrix}x\cr
      y\end{pmatrix}\right ] \; \cr&\quad\times e^{ -px +iqy}\;
  \exp\left \{ -\frac{t}{2} \left[\left(1+\frac{1}{N}-\frac{2}{N^2}
      \right) q + \left(1-\frac{1}{N}-\frac{2}{N^2} \right) p
    \right]\right \}\,.
\end{align}
Using (\ref{identity}) we now perform the sums over $p$ and $q$,
\begin{align}
  \bar R(v) &= -\frac{N^2}{t}\;e^{-\frac{t}{2}
    \left(1-\frac{1}{N^2}\right)}\cr
  &\quad\times\int_{-\infty}^{\infty} \int_{-\infty}^{\infty} \frac{dx
    dy}{2\pi} \; \exp\left [ -\frac{N}{2t} [ (1-1/N)x^2 + (1+1/N)y^2
    -2ixy/N ]\right ]\cr&\quad\times \int_0^1 d\rho \, \frac { \bigl[
    1-v\rho e^{-x-(t/2) (1-1/N -2/N^2 )} \bigr]^{N-1}}{ \bigl [
    1-v\rho e^{iy-(t/2) (1+1/N -2/N^2 )}\bigr ]^{N+1}}\,.
  \label{rv}
\end{align}
Note that because of $|v|<1$ the denominator in the last line is never
zero.  The integral over $\rho$ can be done exactly, if one wishes,
resulting in
\begin{align}
  \bar R(v) &= \frac{N}{t}\;e^{-\frac{t}{2}
    (1-\frac{1}{N^2})}\;\int_{-\infty}^{\infty}
  \int_{-\infty}^{\infty} \frac{dx dy}{2\pi} \left \{ \left [\frac {
        1-v\, e^{-x-(t/2) (1-1/N -2/N^2 )}}{1-v\, e^{iy-(t/2) (1+1/N
          -2/N^2 )}}\right ]^N-1\right\} \; \cr 
  &\quad\times \frac{e^{ -\frac{N}{2t} [ (1-1/N)x^2 + (1+1/N)y^2
      -2ixy/N ]}}{v\left [ e^{-x-(t/2)(1-1/N-2/N^2)} -
      e^{iy-(t/2)(1+1/N-2/N^2)}\right ] }\,.
\end{align}
The above formula was derived for $|v|<1$; this is enough for finding
$\rho^{\tr}_N (\theta )$ via \eqref{eq:rhotrue}. Using symmetries of
$\langle R(u,v,W)\rangle $ one can immediately write down also results
for $|v| > 1$.

\subsection[Making sense of negative integer $N$]{\boldmath Making
  sense of negative integer $N$}
\label{sec:neg}

Conforming to previous observations (see~\cite{Dunne:1988ih} and
references therein), we extend our result to negative integer $N$.
This may be of relevance to $1/2N$ playing the role of the viscosity
term in Burgers' equation~\cite{Neuberger:2008mk,Neuberger:2008ti} and
also to approximate equations in~\cite{Blaizot:2009kj}.

We first restate the result derived earlier,
\begin{align}
  \bar R(u,v,N)&\equiv\braket{R(u,v,W)}\notag\\
  &=1+\frac{u+v}{N} \sum_{p=0}^{N-1}\sum_{q=0}^\infty \frac{1}{p+q+1}
  u^p v^q e^{-\frac{\lambda {\cal A}}{2} {\hat C}(p,q,N)} M^A (p,N)
  M^S (q, N)\,,
\label{R}
\end{align}
where
\begin{subequations}\label{factors}
  \begin{align}
    {\hat C} (p,q,N) &= \frac{C(p,q,N)}{N} 
    =(p+q+1)\left ( 1-\frac{p+q+1}{N^2} +\frac{q-p}{N}\right )\,,\\
    M^A(p,N)&=\frac{(N-p)(N-p+1)\cdots N}{(p+1)!} (p+1)\,,\\
    M^S (q,N)&=\frac{(N+q)(N+q-1)\cdots N}{(1+q)!} (q+1)\,.
  \end{align}
\end{subequations}
In equations~(\ref{factors}) $p$ and $q$ still are nonnegative integers, 
but $N$ is allowed to be an integer of arbitrary sign (with
$N=0$ excluded). 

Note that for $p\ge N$, $M^A(p,N)=0$. Hence, still keeping $N>0$, we
can remove one of the restrictions on the range of $p$ in the sum in
equation~\eqref{R},
\begin{equation}
  \bar R(u,v,N)=1+\frac{u+v}{N} \sum_{p,q=0}^\infty
  \frac{1}{p+q+1} u^p v^q e^{-\frac{\lambda {\cal A}}{2} {\hat C}(p,q,N)}
  M^A (p,N) M^S (q, N)\,.
\label{loca}
\end{equation}
Observe
\begin{subequations}\label{signa}
  \begin{align}
    {\hat C} (p,q,-N)&= {\hat C} (q,p,N)\,,\\
    M^A(p,-N)&=(-1)^{p+1} M^S (p,N)\,,\\
    M^S(q,-N)&=(-1)^{q+1} M^A(q,N)\,.
  \end{align}
\end{subequations}
The entire dependence on $N$ in~(\ref{loca}) is explicit, and the
function $\bar R(u,v,N)$ remains well-defined for $N<0$, so long as
the fixed parameter $\lambda{\cal A}$ is positive.  With $N>0$ this
leads to
\begin{equation}
  \bar R(u,v,N)=1+\frac{-u-v}{-N} 
  \sum_{p,q=0}^\infty
  \frac{(-u)^p (-v)^q }{p+q+1}M^S (p,-N) M^A (q,-N)
  e^{-\frac{\lambda{\cal A}}{2} {\hat C}(q,p,-N)}\,.
\end{equation}
Interchanging the dummy summation labels $p$ and $q$ we get
\begin{equation}
  \bar R(u,v,N)=\bar R(-v,-u,-N)\,.
\end{equation}
Writing
\begin{equation}
  \bar R(u,v,N)=1+\frac{u+v}{N} \Omega(u,v,N)
\end{equation}
produces
\begin{equation}
  \Omega(u,v,N)=\Omega(-v,-u,-N)\,.
\end{equation}
Now set $u=-v$. $\Omega(-v,v,N)$ is finite for $\lambda{\cal A}>0$. We
have
\begin{equation}
  \label{eq:Osym}
  \Omega(-v,v,N)=\Omega(-v,v,-N)\,.
\end{equation}
$\Omega(-v,v,N)$ determines $\rho^{\tr}_N (\theta,t)$ via
\eqref{eq:rhotrue} because of $\Omega(-v,v,N)=N\bar R(v)$, i.e.,
\begin{align}
  \rho^{\tr}_N (\theta,t)=1+\frac2{N^2}\lim_{\epsilon\to 0^+}
  \re\,[v\, \Omega(-v,v,N)]\,,\qquad v=e^{-\epsilon+i\theta}\,.
\end{align}
At this point we realize that we have defined $\rho^{\tr}_N
(\theta,t)$ for negative integer $N$, too:
\begin{align}
  \rho^{\tr}_{-N}(\theta, t)&=1 + \frac{2}{N} \lim_{\epsilon\to 0^+}
  \re\,[v\, \Omega(-v,v,-N)]
  =\rho^{\tr}_N (\theta,t)\,,
\end{align}
where in the last step we have observed \eqref{eq:Osym}.

\subsection[Large-$N$ asymptotics]{\boldmath Large-$N$ asymptotics}

If one could expand $\rho^{\tr}_N (\theta, t)$ in $N$ around $N=0$,
only even powers of $N$ would enter.  However, all one can do is an
asymptotic expansion in $1/N$, and then odd powers will appear.
Essentially, the asymptotic expansion is not in $1/N$ but rather in
$1/|N|$.  For example, for small loops there is an arc centered at
$\pm \pi$ where the infinite-$N$ eigenvalue density has a gap, and
there at finite $N$ one has exponential suppression of the form
$\exp(-|N|\kappa)$, $\kappa >0$ --- it makes no sense to drop the
absolute value on $N$ in the exponent.  As another example, consider a
sub-leading term that goes like $\cos (N\theta)/|N|$.  The oscillatory
behavior of $\rho^{\tr}_N (\theta, t)$ comes from a contribution of
this type.

We now turn to the integral representation to take the first steps in
a $1/N$ expansion of $\rho^{\tr}_N (\theta,t)$.  Shifting integration
variables $x\to x+(t/2)\left(1/N+2/N^2\right)$ and $y\to
y-i(t/2)\left(1/N-2/N^2\right)$ in~(\ref{rv}), we obtain
\begin{equation}
  \label{Rint}
  \bar R(v)=-\frac{N^2}{t}e^{-\frac t2}\int\!\!\int_{-\infty}^\infty
  \frac{dxdy}{2\pi}\int_0^1d\rho\, 
  e^{-\frac N{2t}(x^2+y^2)+\frac1{2t}(x+iy)^2-\frac12(x-iy)} \frac{\left[1-v\rho e^{-x-t/2}\right]^{N-1}}{\left[1-v\rho e^{iy-t/2}\right]^{N+1}}\,.
\end{equation} 
Since this integral representation was derived for $|v|<1$, we set
$v=e^{i \theta-\epsilon}$ with $|\theta|\leq\pi$, $\epsilon>0$, and
take the limit $\epsilon\to 0$ at the end.  We write~(\ref{Rint}) as
\begin{align}
  \label{RintLog}
  \bar R(v)&=-\frac{N^2}{t}e^{-\frac t2}\int\!\!\int_{-\infty}^\infty
  \frac{dxdy}{2\pi}\int_0^1d\rho\, e^{-\frac
    N{2t}\left(x^2+y^2\right)+\frac1{2t}(x+iy)^2-\frac12(x-iy)}\cr
  &\quad\times e^{(N-1)\log\left(1-v\rho
      e^{-x-t/2}\right)-(N+1)\log\left(1-v\rho e^{iy-t/2}\right)}\,.
\end{align}
At large $N$, the integrals over $x$ and $y$ decouple at leading order
and can be done independently by saddle-point approximations.  Let us
start with the integral over $y$ since it is conceptually simpler. The
$y$-dependent coefficient of the term in the exponent in
equation~(\ref{RintLog}) that is proportional to $-N$ is
\begin{equation}
  \bar f(y)=\frac1{2t} y^2+\log\left[1-v\rho e^{iy-\frac t2}\right]\,.
\end{equation}
Substituting $y=u-it/2=it(U-1/2)$ (with $u=itU$ in analogy to
section~\ref{secSymmSaddle}) results in exactly the same integrand that
was already considered in section~\ref{secSymmSaddle}, with the
replacements $T\to t$ and $z\to1/v\rho$ (with $|v\rho|<1$) and with an
integration over $u$ that is now along the line from $-\infty+it/2$ to
$+\infty+it/2$.  Since there are no singularities between this line
and the real-$u$ axis we can change the integration path to be along
the real-$u$ (or imaginary-$U$) axis.  Now everything goes through as
in section~\ref{secSymmSaddle}.  The saddle-point equation reads
\begin{equation}
  \label{SaddleU}
  e^{-tU}\frac{U+1/2}{U-1/2}=\frac1{v\rho}\,,
\end{equation}
which is equivalent to \eqref{contour}.  In figure~\ref{figSaddle} we
show the contours in the complex-$U$ plane on which the solutions of
the saddle-point equation have to lie. (For sufficiently small $\rho$
we now encounter the case mentioned in section~\ref{sec:spa} where for
$t>4$ the closed contour in the left half-plane is missing.)  The
relevant saddle point, which we denote by $y_0(\theta,t,\rho)$, is
again on the closed contour in the right half-plane.  For decreasing
$\rho$ this contour contracts, but this makes no difference to our
analysis.  The result for the $y$-integral is given by an expression
similar to \eqref{eq:spgauss}.

\FIGURE[t]{
  \includegraphics[width=0.3\textwidth]{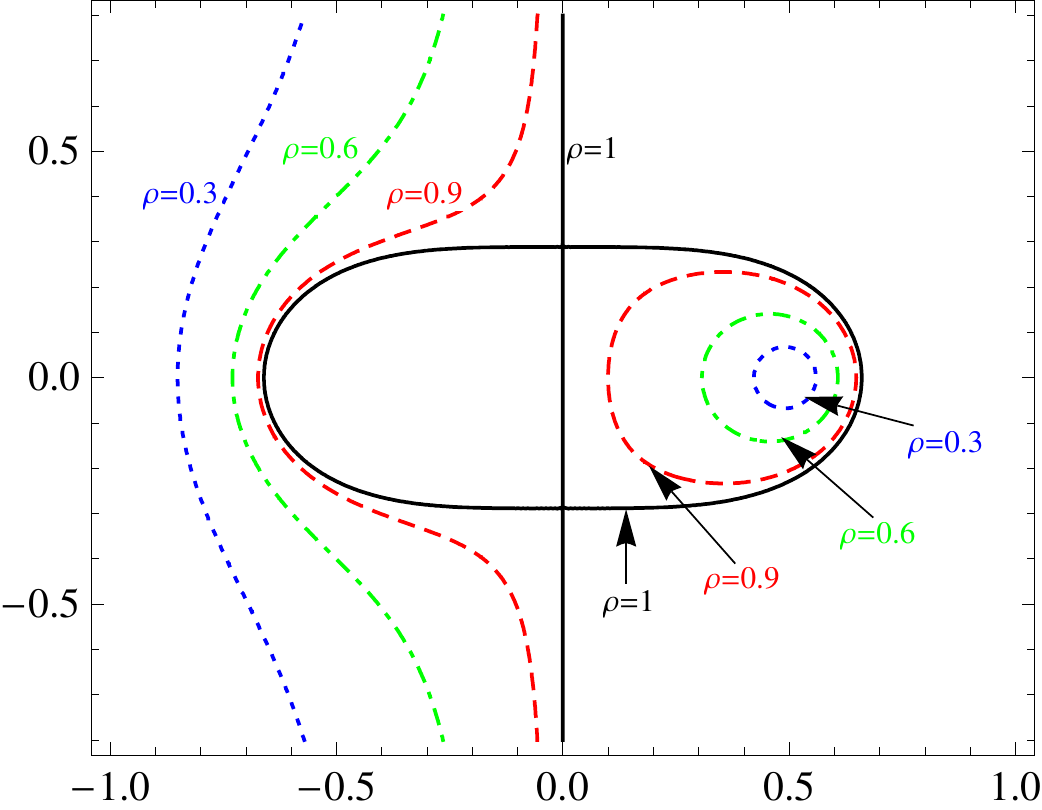}\hfill
  \includegraphics[width=0.3\textwidth]{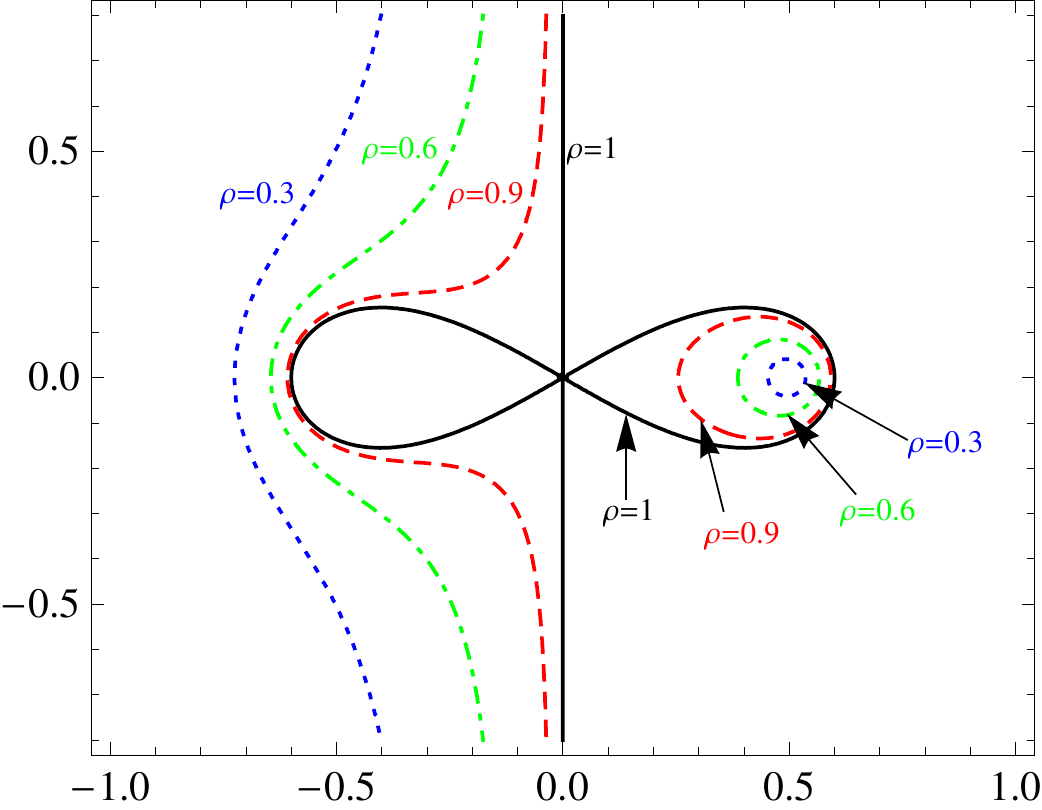}\hfill
  \includegraphics[width=0.3\textwidth]{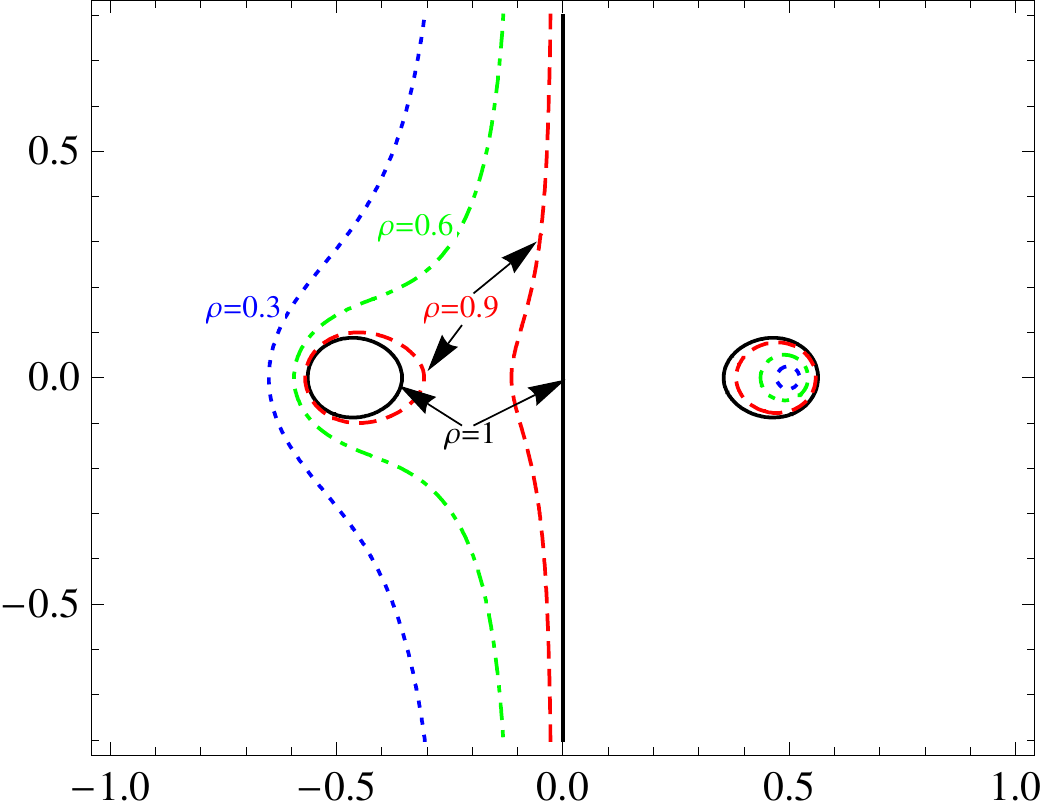}
  \caption{Contours of solutions of equation~({\protect \ref{SaddleU}}) in
    the complex-$U$ plane at $t=3$ (left), $t=4$ (middle), and $t=5$
    (right) for $\rho=1$ (black, solid), $\rho=0.9$ (red, dashed),
    $\rho=0.6$ (green, dot-dashed), and $\rho=0.3$ (blue, dotted).  In
    the figures (but not in the analysis) we have taken $|v|=1$ for
    simplicity.}
  \label{figSaddle}
}

We now turn to the integral over $x$.  The $x$-dependent coefficient
of the term in the exponent in equation~(\ref{RintLog}) that is
proportional to $-N$ is
\begin{equation}
  \tilde f(x)=\frac1{2t} x^2-\log\left[1-v\rho
    e^{-x-t/2}\right]=-\bar f(ix)\,. 
\end{equation}
Substituting $x=-iu-t/2=t(U-1/2)$ (with $u=itU$) again leads to the
integral considered in section~\ref{secSymmSaddle} and the
saddle-point equation \eqref{SaddleU}, except that the integration is
now along the real-$U$ axis.  The positions of the saddle points of
the $x$-integral are obtained by rotating the saddles of the
$y$-integral by $-\pi/2$ in the complex-$U$ plane, i.e., $x_s=-iy_s$.
At a saddle point we have
\begin{equation}
  \label{fppx0}
  \tilde f''(x_s)=\frac 1t+\frac{x_s}t\left(1+\frac
    {x_s}t\right)=\bar f''(y_s) \,,
\end{equation}
and therefore the directions of steepest descent through a saddle
$y_s$ and the corresponding saddle $x_s=-iy_s$ are identical (no
rotation).  By analyzing the directions along which the phase of the
integrand is constant, we find that the integration contour can always
be deformed to go through the (single) saddle-point in the right
half-plane in the direction of steepest descent.  Depending on the
parameters $\rho$, $v$, and $t$, there is either one or no additional
saddle point on the contour(s) in the left half-plane through which we
can also go in the direction of steepest descent.  If there is such an
additional saddle point, we find that its contribution to the integral
is always exponentially suppressed in $N$ compared to the saddle point
in the right half-plane and can therefore be dropped from the
saddle-point analysis.  In addition, there are infinitely many more
saddle points on the open contour in the left half-plane.  However, we
cannot deform the integration path to go through these points in the
direction of steepest descent and therefore do not need to include
them. An example for the location of the saddle points and the
deformation of the integration path is given in
figure~\ref{fig:profile}.  To summarize, the $x$-integral can be
approximated by the contribution of the single saddle point in the
right half-plane, which again leads to an expression similar to
\eqref{eq:spgauss}.

\FIGURE[t]{
  \centerline{\includegraphics[width=100mm]{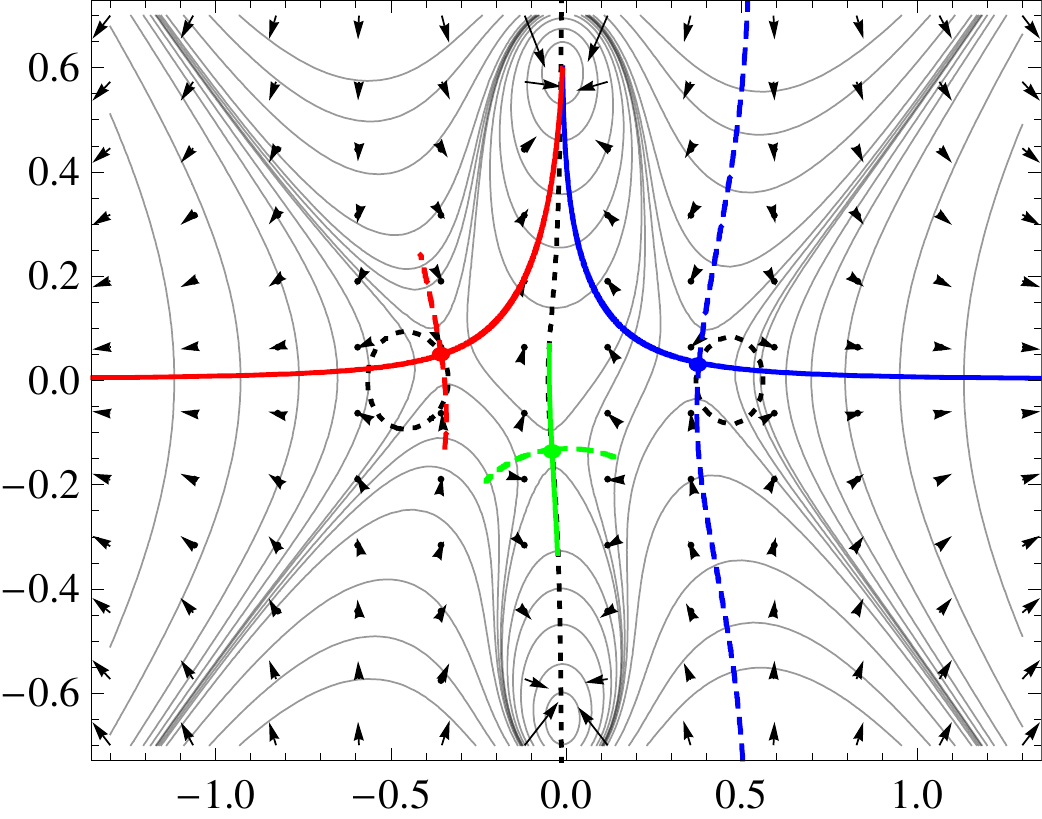}}
  \caption{Example for the location of the saddle points and the
    deformation of the integration path in the complex-$U$ plane for
    $t=5$ and $\rho=0.95$.  The dashed black curves (two closed, one
    open) are the curves on which all saddle points have to lie,
    cf.~\eqref{eq:uiur}.  In this example $\theta=3.0$.  On each of
    the closed curves there is one saddle point (red dot and blue
    dot), and on the open curve there are infinitely many saddle
    points, but only one of them in the region shown in the plot
    (green dot).  The thin solid lines are lines of constant $\re
    \tilde f(x)$ and $\re \bar f(y)$.  The arrows point in the
    direction of increasing $\re \tilde f(x)$ or decreasing $\re \bar
    f(y)$.  The dashed blue curve is the integration path for the
    $y$-integral along the direction of steepest descent.  The solid
    red-blue curve is the integration path for the $x$-integral along
    the direction of steepest descent.}
  \label{fig:profile}
}

Combining the saddle-point approximations for the integrals over $x$
and $y$, we find that, up to exponentially small corrections in $N$,
the integral in equation~(\ref{RintLog}) is given by
\begin{equation}
  \bar R(v)=-\frac{N^2}{t}e^{-t/2}\int_0^1d\rho\,
  \frac1{2\pi}\left(\frac{2\pi}{N\tilde f''(x_0)}\right)
  \frac1{(1-v\rho e^{-x_0-t/2})^2}\,e^{-x_0}\,, 
\end{equation}
where $x_0=x_0(\theta,t,\rho)$ is the dominating saddle point of the
$x$-integral. $x_0$ is a solution of the saddle-point equation
obtained by differentiating $\tilde f(x)$, which can be written as
\begin{equation}
  \label{Saddlex0}
  v\rho e^{-x_0-t/2}=\frac{x_0}{x_0+t}
\end{equation}
and leads to
\begin{equation}
  \left(1-v\rho e^{-x_0-\frac t2}\right)^2=\left(\frac t{t+x_0}\right)^2\,.
\end{equation}
With~(\ref{fppx0}) we obtain
\begin{equation}
  \tilde f^{\prime\prime}(x_0)\left(1-v\rho e^{-x_0-\frac
      t2}\right)^2=\frac{t+x_0\left(t+x_0\right)}{\left(t+x_0\right)^2} 
\end{equation}
and
\begin{equation}
  \bar R(v)=-\frac Nt e^{-\frac t2}\int_0^1d\rho\,
  \frac{\left(t+x_0\right)^2}{t+x_0\left(t+x_0\right)}\,e^{-x_0}\,.  
\end{equation}
Differentiating equation~(\ref{Saddlex0}) with respect to $\rho$ leads to
\begin{align}
  \frac{\partial x_0}{\partial\rho}&=\frac1\rho
  \frac{x_0\left(t+x_0\right)}{t+x_0\left(t+x_0\right)}
  =ve^{-x_0-t/2}\frac{\left(t+x_0\right)^2}{t+x_0\left(t+x_0\right)}\,,
\end{align}
which yields
\begin{equation}
  \bar R(v)=-\frac{N}{tv}\int_0^1d\rho\, \frac{\partial x_0}{\partial \rho}
  =-\frac{N}{tv}\left[x_0(\theta,t,\rho=1)-x_0(\theta,t,\rho=0)\right]\,.
\end{equation}
We know from \eqref{Saddlex0} that $x_0(\theta,t,\rho=0)=0$. If we
parametrize $x_0(\theta,t,\rho=1)=\lambda(\theta,t) t$, where
$\lambda(\theta,t)$ has to solve
\begin{equation}
  \label{SaddleLambda}
  \lambda=\frac{1}{\frac1v e^{t\left(\lambda+1/2\right)}-1}\,,
\end{equation}
and take the limit $\epsilon\to0^+$, we end up with
\begin{equation}
\bar R(v)=-\frac{N\lambda(\theta,t)}v\,,\qquad v=e^{i\theta}\,.
\end{equation}
Here we need to keep in mind that we have to pick the solution of
equation~(\ref{SaddleLambda}) which corresponds to the dominating saddle
point $x_0$ of the $x$-integral for $|v\rho|<1$.

Using \eqref{eq:rhotrue} we obtain
\begin{align}
  \lim_{N\to\infty}\rho_N^\tr(\theta,t)=1+2\re\lambda(\theta,t)\,,
\end{align}
which is equal to $\rho_\infty(\theta,t)$
\cite{Durhuus:1980nb,Janik:2004tw}.  Keeping higher orders in the
saddle-point approximation (as explained in section~\ref{sec:spa1N}),
we can compute the asymptotic expansion of $\rho_N^\tr(\theta,t)$ in
powers of $1/N$.

\subsection{A partial differential equation for the average of the
  ratio of characteristic polynomials at different arguments}

In the expression for $\Omega (u,v,N)$ that follows from~(\ref{loca})
a derivative with respect to $t$ will bring down the Casimir factor
from the exponent. Writing
\begin{equation}
  u=-e^{X+Y}\,,\qquad v=e^{X-Y}\,,\qquad
  f_N(X,Y,t)=\Omega(u,v,N)|_{t=\lambda{\cal A}}
\end{equation}
we can reconstruct the Casimir by derivatives with respect to $X$ and
$Y$. All that comes in is the bilinear structure of the Casimir. We
obtain
\begin{equation}
  \frac{\partial f_N}{\partial t}=\frac{1}{2}\left [
    \frac{1}{N^2}\left(\frac{\partial}{\partial X}+1\right )^2 -\left
      (1-\frac{1}{N}\frac{\partial}{\partial Y}\right ) \left
      (\frac{\partial}{\partial X}+1\right ) \right ] f_N\,. 
\end{equation}
One can simplify the equation by $f_N\to g_N=e^{X-NY} f_N$,
\begin{equation}
  \frac{\partial g_N}{\partial t}=\frac{1}{2}\left (
    \frac{1}{N^2}\frac{\partial^2}{\partial X^2}
    +\frac{1}{N}\frac{\partial^2}{\partial Y \partial X} \right )
  g_N\,. 
\end{equation}
Rescaling $X\to N X=Z$ removes all explicit dependence on $N$ in the
equation. The equation is linear, so we are free to rescale $g_N$ by
any power of $N$ we find convenient.  We define
\begin{equation}
  G_N(Z,Y,t)\equiv \frac{1}{N} g_N(Z/N,Y,t)
\end{equation}
and now have
\begin{equation}
  \frac{\partial G_N}{\partial t}=\frac{1}{2}\left (
    \frac{\partial^2}{\partial Z^2}  +\frac{\partial^2}{\partial
      Y \partial Z} \right ) G_N\,. 
\label{pde}
\end{equation}
The $N$-dependence of $G_N$ will then come in only through the initial
condition at $t=0$.  We proceed to find the initial condition.
Similarly to~(\ref{identity}) the combinatorial factors $M^{A,S}$ have
the following generating functions:
\begin{subequations}
  \begin{align}
    \sum_{p=0}^\infty M^A (p,N) A^p &= N(1+A)^{N-1}\,,\\
    \sum_{q=0}^\infty M^S (q,N) S^q &= \frac{N}{(1-S)^{N+1}}\,.
  \end{align}
\end{subequations}
These identities go beyond~(\ref{identity}) in that they hold also for 
negative integer $N$. Using 
\begin{equation}
  \frac{1}{p+q+1} =\int_0^1 d\rho\, \rho^{p+q}
\end{equation} 
and the fact that at $t=0$ we have
\begin{equation}
  \Omega(u,v,N)|_{t=0} =\sum_{p,q=0}^\infty 
  \frac{u^p v^q}{p+q+1} M^A (p,N) M^S (q, N)\,
\end{equation}
we obtain
\begin{equation}
  \Omega(u,v,N)|_{t=0} =N^2 \int_0^1 d\rho \,
  \frac{(1+\rho u)^{N-1}}{(1-\rho v )^{N+1}}\,.
\end{equation}
Observing that
\begin{equation}
  \frac{\partial }{\partial r } \frac{(1+r A)^N}{(1+rB)^N}=
  N(A-B)\frac{(1+rA)^{N-1}}{(1+rB)^{N+1}}
\end{equation}
we derive
\begin{equation}
  \Omega (u,v,N)|_{t=0} = \frac{N}{u+v} 
  \left [ \left (\frac{1+u}{1-v}\right )^N -1\right ]\,.
\end{equation}
From this we now find the initial condition associated with
equation~(\ref{pde}),
\begin{equation}
  \label{eq:Ginit}
  G_N(Z,Y,t=0) = -\frac{e^{-NY}}{e^{Y}-e^{-Y}} 
  \left [ \left (
      \frac{1-e^{\frac{Z}{N}+Y}}{1-e^{\frac{Z}{N}-Y}}\right )^N -1
  \right ]\,. 
\end{equation}
The partial differential equation \eqref{pde} and the associated
initial condition \eqref{eq:Ginit} admit arbitrary $N$, no longer
restricted to integers, although for noninteger $N$ periodicity in
$\theta$ is lost.  However, periodicity in $\theta$ was assumed when
the relation between $\rho^{\tr}_N$ and $\Omega$ was derived.

One can again check whether there is a symmetry under $N\to -N$. The
partial differential equation is linear and invariant under
\begin{equation}
  Z\to -Z\,,\quad Y\to -Y\,,\quad N\to -N\,.
\end{equation}
The initial condition switches sign under this transformation.  Hence,
\begin{equation}
  \label{eq:Gsym}
  G_N (Z,Y,t)= - G_{-N} (-Z,-Y,t)\,.
\end{equation}
For noninteger $N$ there is some subtlety in defining the cuts in the
initial condition so that the above holds.

By Fourier/Laplace transforms one can derive integral representations,
embedding the initial condition at $t\to 0$. To get to the density
$\rho^{\tr}_N(\theta,t) $ via \eqref{eq:rhotrue} and $\bar
R(v)=\Omega(-v,v,N)/N$, one needs to set $u=-v$, which corresponds to
$Y=0$ at fixed $Z/N=-\epsilon+ i\theta$, i.e.,
\begin{equation}
  \label{eq:rhoG}
  \rho^{\tr}_N (\theta,t) =1 +\frac{2}{N} \lim_{\epsilon\to 0^+} \re\,
  G_N(N(-\epsilon+i\theta),0,t)\,. 
\end{equation}

At $t>0$ the limit should be smooth, but at $t=0$ one needs to
generate a $\delta$-function singularity in $\rho^{\tr}_N(\theta,t)$
at $\theta=0\bmod 2\pi$.  We first need the $Y\to0$ limit of
\eqref{eq:Ginit}, which is
\begin{align}
  G_N(Z,Y=0,t=0)=\frac {Ne^{Z/N}}{1-e^{Z/N}}
  =\frac {Ne^{-\epsilon+i\theta}}{1-e^{-\epsilon+i\theta}}\,. 
\end{align}
Expanding the denominator in a geometric series and using
\eqref{eq:rhoG} yields
\begin{align}
  \rho^{\tr}_N (\theta, t=0)=1+e^{i\theta}\sum_{k=0}^\infty
  e^{ik\theta}+e^{-i\theta}\sum_{k=0}^\infty e^{-ik\theta}
  =\sum_{k=-\infty}^\infty e^{ik\theta}
  =2\pi\delta_{2\pi}(\theta)
\end{align}
as expected.  $\rho^{\tr}_N(\theta, t=0)$ is independent of $N$. 

\section{Comparison of the three eigenvalue densities}
\label{sec:compare}

\subsection[$\rho_N^{\tr}(\theta, t)$ and
$\rho_N^{\sy}(\theta,T)$]{\boldmath $\rho_N^{\tr}(\theta, t)$ and
  $\rho_N^{\sy}(\theta,T)$}

If we want to compare $\rho_N^{\tr}(\theta,t)$ and
$\rho_N^{\sy}(\theta,T)$ we have to take into account the $1/N$
difference between $t$ and $T$, see equation~(\ref{defT}). At fixed
$N$ and $t$, we have to compare $\rho_N^{\tr}(\theta,t)$ and
$\rho_N^{\sy}(\theta,T=t(1-1/N))$. The densities $\rho_N^{\tr}$ and
$\rho_N^{\sy}$ can be obtained numerically by evaluating the sums in
equation~(\ref{barRsum}) and equation~(\ref{rhosym}), respectively.

\FIGURE[t]{
  \begin{tabular}{l@{\hspace*{5mm}}r}
    \includegraphics[width=0.47\textwidth]{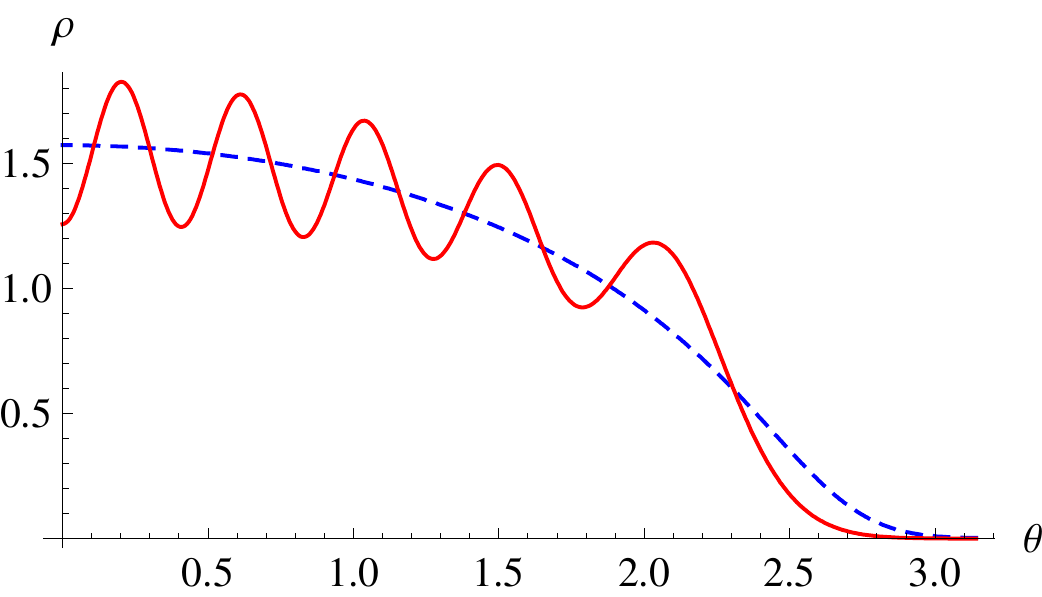} &
    \includegraphics[width=0.47\textwidth]{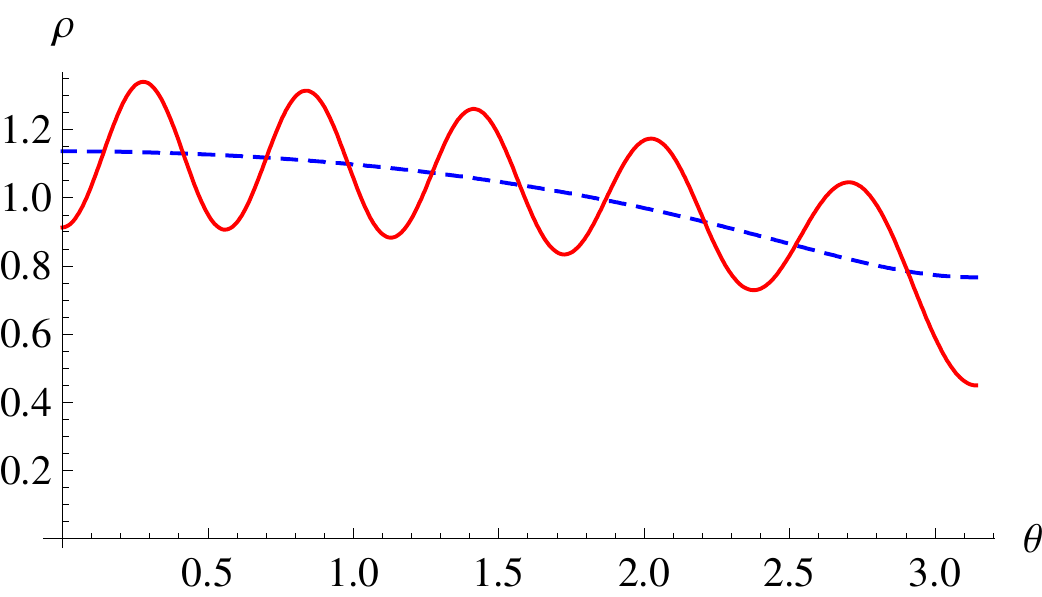} \\
    \includegraphics[width=0.47\textwidth]{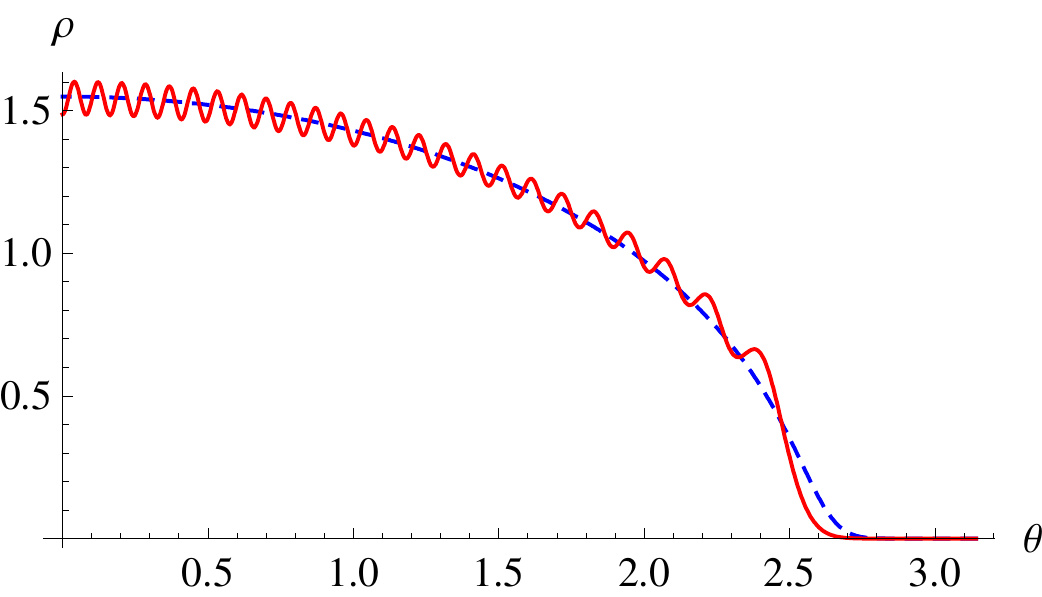} &
    \includegraphics[width=0.47\textwidth]{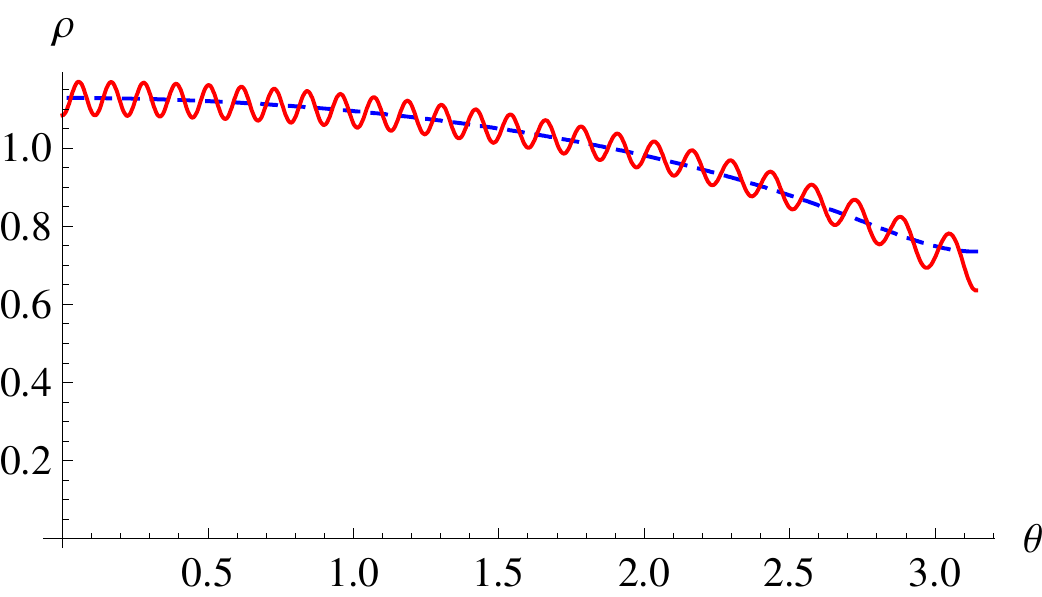} 
  \end{tabular}
  \caption{Plots of the densities $\rho_N^\tr(\theta,t)$ (red, solid)
    and $\rho_N^\sy(\theta,T)$ (blue, dashed) for $t=2$ (left) and
    $t=5$ (right), $N=10$ (top), and $N=50$ (bottom).}
  \label{figRhoTrueSymm}
}

Figure~\ref{figRhoTrueSymm} shows plots of
$\rho_N^\tr(\theta,t)=\rho_N^\tr(-\theta,t)$ and
$\rho_N^\sy(\theta,T)=\rho_N^\sy(-\theta,T)$ for $t=2$, $t=5$, $N=10$,
and $N=50$ in the interval $0\le\theta\le\pi$.  As stated in
section~\ref{subsecRhoSymm}, $\rho_N^\sy(\theta,T)$ decreases
monotonically in that interval. The true eigenvalue density
$\rho_N^\tr(\theta,t)$ has $N$ peaks (in the complete interval
$[-\pi,\pi]$) and oscillates around the nonoscillatory function
$\rho_N^\sy(\theta,T)$.

\subsection[$\rho_N^{\tr}(\theta, t)$ and $\rho_N^\as(\theta,
\tau)$]{\boldmath $\rho_N^{\tr}(\theta, t)$ and $\rho_N^\as(\theta,
  \tau)$}

The density $\rho_N^\as(\theta,\tau)$ is given by a sum of $N$
$\delta$-functions, located at the zeros of the average characteristic
polynomial, see
section~\ref{subsecRhoAsym}. Figure~\ref{figRhoTrueZeros} shows that
the locations of these zeros are close to the positions of the $N$
peaks of $\rho_N^\tr(\theta,t)$. Here we again have to take into
account the $1/N$ difference in the definitions of $t$ and $\tau$. For
fixed $N$ and $t$, the peaks of $\rho_N^\tr(\theta,t)$ have to be
compared to the zeros of $\la\det(e^{i\theta}-W)\ra$ at
$\tau=t(1+1/N)$.

\FIGURE[t]{
  \begin{tabular}{l@{\hspace*{5mm}}r}
    \includegraphics[width=0.47\textwidth]{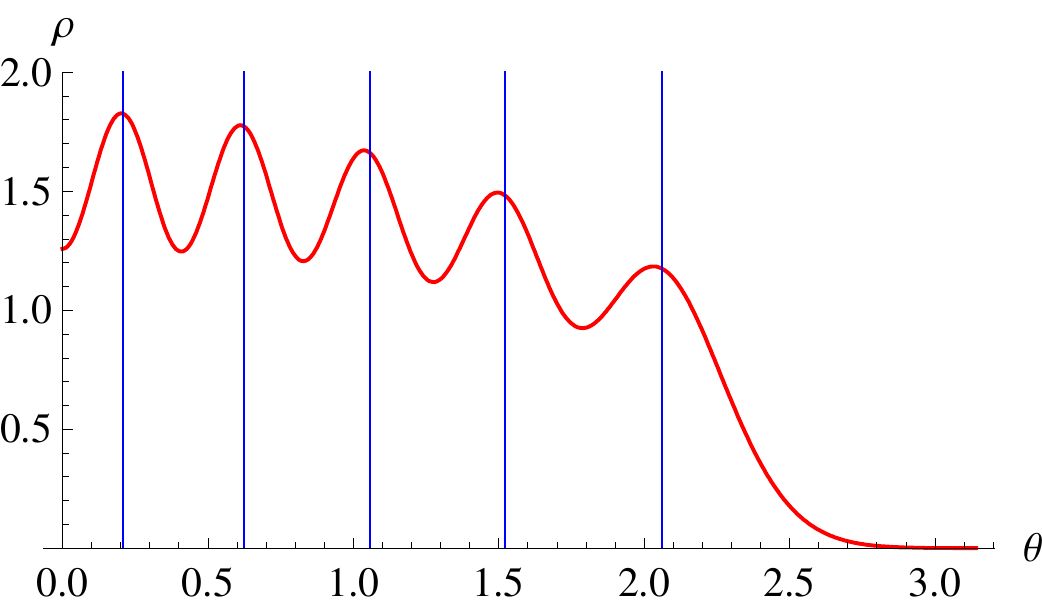} &
    \includegraphics[width=0.47\textwidth]{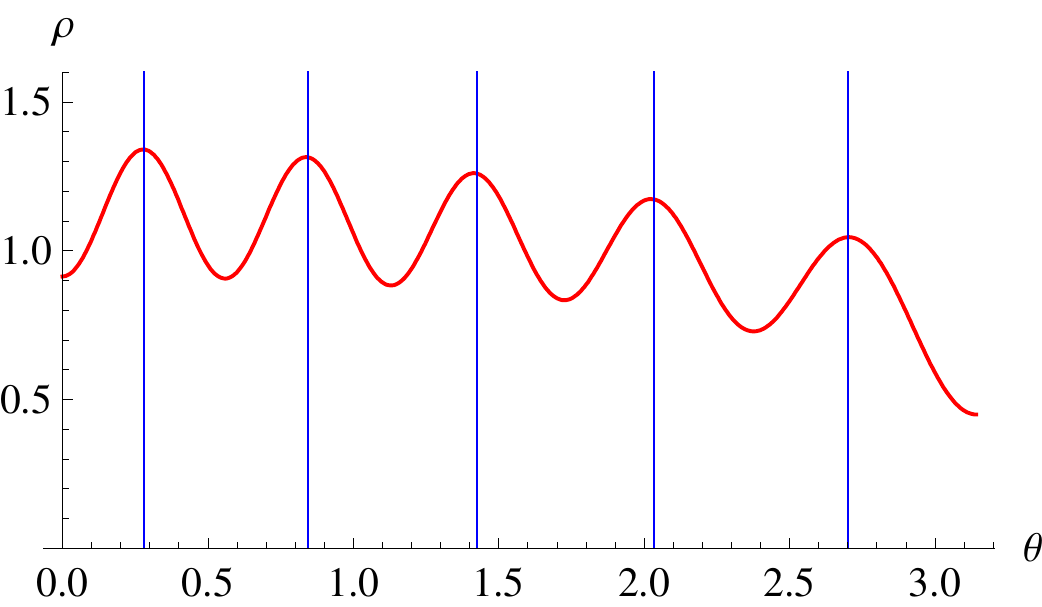} \\
    \includegraphics[width=0.47\textwidth]{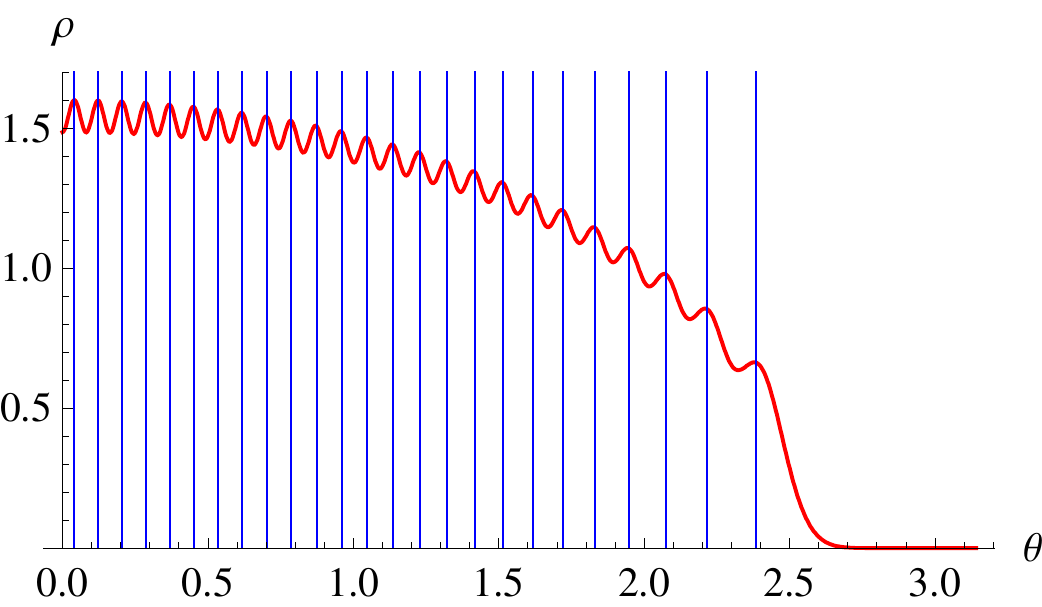} &
    \includegraphics[width=0.47\textwidth]{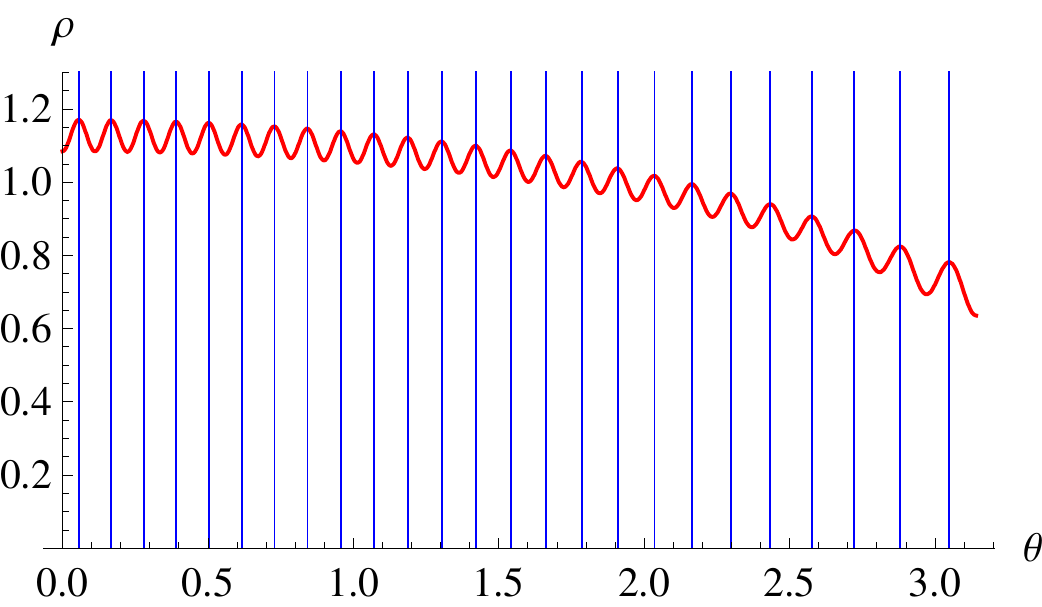} 
  \end{tabular}
  \caption{Plots of the density $\rho_N^\tr(\theta,t)$ (oscillatory
    red curve) together with the positions of the zeros of
    $\la\det(e^{i\theta}-W)\ra$ (vertical blue lines) for $t=2$ (left)
    and $t=5$ (right), $N=10$ (top), and $N=50$~(bottom).}
  \label{figRhoTrueZeros}
}

Computing the positions of the peaks and valleys of $\rho^\tr$ and the
corresponding zeros of the average characteristic polynomial for large
$N$ shows that the difference in position between a peak and its
matching zero vanishes faster than the difference in position between
that peak and the next valley.  This means that
\begin{equation}
 \gamma=\left|\frac{\theta^\text{(peak)}-\theta^\text{(matching
       zero)}}{\theta^\text{(peak)}-\theta^\text{(next\
       valley)}}\right|  
\end{equation}
scales like
\begin{equation}
  \gamma\propto N^{-\mu}\quad\text{with}\quad\mu>0\,.
\end{equation}
It turns out that the value of the exponent $\mu$ depends on $t$ and
may be different in different parts of the spectrum, but it is always
positive (for large $N$).

In the bulk of the spectrum, the difference between peak and
neighboring valley scales like $N^{-1}$, whereas the difference
between peak and matching zero scales like $N^{-2}$ for all $t$. This
results in $\mu_{\rm bulk}=1$.  Figure~\ref{figPeakZeroPos} shows a
plot of $\log\gamma$, computed for the peak closest to $\theta=0$, as
a function of $\log N$ for $t=5$. The line fitted through the data
points has a slope of $-1+O(10^{-3})$. (The reason for choosing
$\theta$ close to $0$ is that stable fit results can be obtained for
lower values of $N$.)
 
For $t>4$, the infinite-$N$ limit of the eigenvalue density,
$\rho_\infty(\theta,t)$, has no gap. In this case the scaling behavior
does not change as one goes to higher $|\theta|$, but it is necessary
to go to large values of $N$ to get stable fit results for $\mu$ when
$|\theta|$ is close to $\pi$. (E.g., for $t=5$ a fit at $N\approx1000$
results in $\mu\approx1.04$ for the extremal peak.)

At the transition point the situation is different. From
equation~(\ref{zMcritical}) we know that the difference between the
position of the extremal zero (the zero closest to $\pi$) and $\pi$
scales like $N^{-0.75}$ for $\tau=4$. Between $N=1800$ and $N=2800$,
the difference between the extremal zero and its critical-$\tau$
approximation scales roughly like $N^{-1.25}$, the difference between
that zero and the extremal peak position scales like $N^{-1.11}$, and
the difference between the positions of the peak and the next valley
(the valley that is closer to $\theta=0$) scales like
$N^{-0.83}$. This results in $\mu_{\rm critical}\approx 0.28$. The
plot of $\log\gamma$ for that case (see figure~\ref{figPeakZeroPos})
indicates that the value of $\mu_{\rm critical}$ might slightly
increase as one goes to even higher values of $N$ (which requires more
computation time).

For $t<4$ there is a gap in the spectrum.  In this case, the exponent
$\mu$ also has different values at the edge and the bulk of the
spectrum, but the variation is not as large as it is at the critical
point. E.g., for $t=3$ a fit between $N=1000$ and $N=1500$ results in
$\mu\approx0.64$ for the extremal peak.  For small $|\theta|$ we again
find $\mu=\mu_{\rm bulk}=1$.  Naturally, we expect the exact values of
the various exponents of $N$ that enter to be rational numbers with
denominators 3 or 4 or 12 (see section~\ref{sec:largest}).

\FIGURE[t]{
  \includegraphics[width=0.47\textwidth]{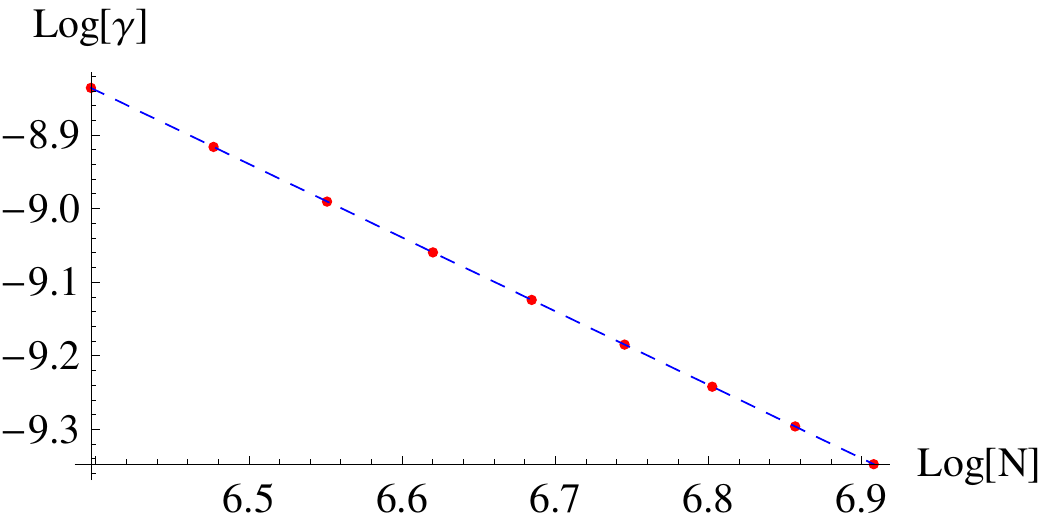}\hfill
  \includegraphics[width=0.47\textwidth]{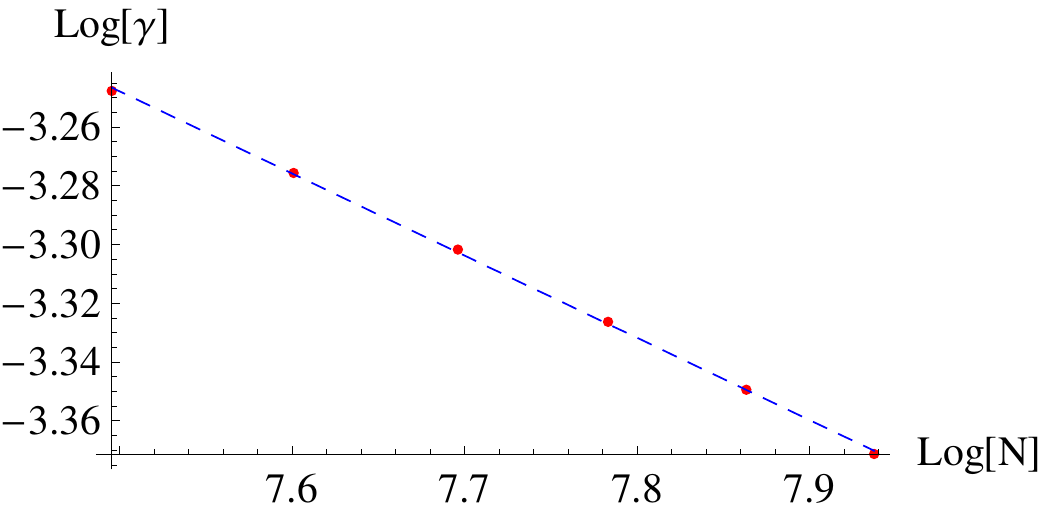}
  \caption{Plots of $\log \gamma$ for the peak closest to $\theta=0$
    at $t=5$ (left) and for the peak closest to $\theta=\pi$ at
    $\tau=4$ (right). Data points (red) are shown together with the
    fitted line (blue, dashed).}
  \label{figPeakZeroPos}
}

\section{The bigger picture}

For definiteness consider four-dimensional (Euclidean) pure $\SU(N)$
Yang-Mills theory. Focus on the string tension and think about a
lattice formulation using a single-plaquette action $S=\sum_p s(U_p)$,
where $p$ denotes a plaquette.  For a fixed $N$ one can choose $s$
such that no phase transitions occur for all bare real couplings
$g_0$.  One can define a ``string tension'', for example, by using a
Creutz ratio,
\begin{equation}
  \sigma_\text{Creutz}(L,g_0)=-\log \frac{\langle \Tr W(L,L+1) \rangle 
    \langle \Tr W(L+1,L) \rangle}{\langle \Tr W(L,L) \rangle 
    \langle \Tr W(L+1,L+1) \rangle }\,,
  \label{creutz}
\end{equation}
where $W(L_1,L_2)$ is the Wilson loop matrix for a rectangular loop
measuring $L_1\times L_2$ in lattice units.

$\sigma_\text{Creutz} (L,g_0 )$ can be expanded around $g_0\to 0$ and
$g_0\to \infty$.  The regimes of validity of these two expansions are
disjoint; in between there is a crossover regime and we can bridge it
only by numerical calculation. There are extra complications around
$g_0=\infty$.  Rectangular loops of the type usually used have a
``roughening'' nonanalyticity in $g_0$.  This nonanalyticity is a
lattice artifact.  It can be avoided by choosing loops at generic
angles with lattice planes.  Then, the definition of
$\sigma_\text{Creutz}$ needs to be extended. All this will increase
the complexity of the strong-coupling expansion. At the end, only a
physical crossover separating the ranges of the weak- and
strong-coupling expansion remains. We have no nonnumerical
calculational method to bridge it.  To get the continuum string
tension in units of the perturbative scale $\Lambda$ we need to take
the continuum limit, a correlated limit in which $g_0\to 0$ and the
overall lattice scale of the loop goes to infinity. This correlated
limit preserves the crossover.

The idea we are pursuing is to improve the above scheme in two
respects.  First, since we wish to set up a calculation in the
continuum we forget about the lattice. Instead of thinking about
$\sigma_\text{Creutz}$ we consider some other observable, for
definiteness the extremal eigenvalue $\theta_M$ of a Wilson loop of
size $\lambda$.

For this to make sense, we need to be able to define $\theta_M$ in
renormalized continuum field theory. We hope that this can be done by
first constructing a renormalized polynomial in $z$ corresponding to
$\langle \det(z - W)\rangle$ and taking the roots of it to define
$\theta_M$.  While we have some idea how a calculation for small loops
might proceed, for large loops we need something beyond ordinary field
theory. Here we assume that an effective string model will describe
$\langle \det(z - W)\rangle$. This model will have a dimensional
parameter, the string tension, and will be a good description for very
large loops, with corrections parametrized by more parameters becoming
more and more important as the loop shrinks.

To relate the string tension to $\Lambda$, the dimensional parameter
entering the perturbation theory for small loops, one needs to join
the two regimes over the crossover.  Here is the point that the
simplification of large $N$ enters: At infinite $N$ the crossover for
$z_M$ collapses into a point, and we have a phase transition. We
postulate that we know that the transition is universal and that we
know it is in the same universality class as the DO transition.

Therefore, for $N\gg1$, the dependence of $z_M$ on intermediate
scales, i.e., scales in the vicinity of the critical scale, is known
up to a few constants.  This is the ingredient that was missing in the
lattice scenario described above. It is now possible to imagine
calculating to some order at short, intermediate, and long scales, and
sew together the three scale ranges. Requiring smooth matches could
produce a number for the string tension in units of the perturbative
scale $\Lambda$.

There are many variations possible. $z_M$ is only one possible example
of a potentially useful variable. $z_M$ depends on the dilation of a
fixed-shaped Wilson loop, measured by a dimensionless variable
$\lambda$.  As a function of $\lambda$, $z_M$ will trace out a
trajectory from $\theta=0$ at $\lambda=0$ (one could replace this by a
$0<\lambda_0\ll 1$) to $\theta=\pi(1-1/2N)$ at $\lambda=\infty$. For
small $\lambda$, the perturbative scale $\Lambda$ enters the
calculations, and for large $\lambda$ the string tension enters.  The
two regimes are joined by the crossover.  To parametrize the crossover
one has to work out the details of the two-dimensional case.

In two-dimensional YM renormalization is trivial, perturbation theory
is well-defined, and there even exists an exact string
description~\cite{Gross:1993hu}. We need to gain control over the
crossover at large $N$, and then we can try to build a prototype of
the calculation we envisage. We also need to learn enough to open the
possibility of finding other interesting observables than $\theta_M$.
This is where the present paper fits in.

\acknowledgments

We acknowledge support by BayEFG (RL), by the DOE under grant number
DE-FG02-01ER41165 at Rutgers University (HN), and by DFG and JSPS
(TW).  HN also notes with regret that his research has for a long time
been deliberately obstructed by his high energy colleagues at Rutgers.
HN thanks G. Dunne for bringing Ref.~\cite{Dunne:1988ih} to his
attention.  TW thanks the Theoretical Hadron Physics Group at Tokyo
University for their hospitality.

\bibliographystyle{JHEP}
\bibliography{tilo} 

\providecommand{\href}[2]{#2}\begingroup\raggedright\begin{thebibliography}{10}

\bibitem{Durhuus:1980nb}
B.~Durhuus and P.~Olesen, {\it {The spectral density for two-dimensional
  continuum QCD}},  {\em Nucl. Phys.} {\bf B184} (1981) 461.

\bibitem{Narayanan:2007dv}
R.~Narayanan and H.~Neuberger, {\it {Universality of large N phase transitions
  in Wilson loop operators in two and three dimensions}},  {\em JHEP} {\bf 12}
  (2007) 066, [\href{http://xxx.lanl.gov/abs/0711.4551}{{\tt
  arXiv:0711.4551}}].

\bibitem{Narayanan:2008he}
R.~Narayanan, H.~Neuberger, and E.~Vicari, {\it {A large N phase transition in
  the continuum two dimensional SU(N) X SU(N) principal chiral model}},  {\em
  JHEP} {\bf 04} (2008) 094, [\href{http://xxx.lanl.gov/abs/0803.3833}{{\tt
  arXiv:0803.3833}}].

\bibitem{Neuberger:2008mk}
H.~Neuberger, {\it {Burgers' equation in 2D SU(N) YM}},  {\em Phys. Lett.} {\bf
  B666} (2008) 106--109, [\href{http://xxx.lanl.gov/abs/0806.0149}{{\tt
  arXiv:0806.0149}}].

\bibitem{Neuberger:2008ti}
H.~Neuberger, {\it {Complex Burgers' equation in 2D SU(N) YM}},  {\em Phys.
  Lett.} {\bf B670} (2008) 235--240,
  [\href{http://xxx.lanl.gov/abs/0809.1238}{{\tt arXiv:0809.1238}}].

\bibitem{Blaizot:2008nc}
J.-P. Blaizot and M.~A. Nowak, {\it {Large $N_c$ confinement and turbulence}},
  {\em Phys. Rev. Lett.} {\bf 101} (2008) 102001,
  [\href{http://xxx.lanl.gov/abs/0801.1859}{{\tt arXiv:0801.1859}}].

\bibitem{GudowskaNowak:2003zx}
E.~Gudowska-Nowak, R.~A. Janik, J.~Jurkiewicz, and M.~A. Nowak, {\it {Infinite
  Products of Large Random Matrices and Matrix-valued Diffusion}},  {\em Nucl.
  Phys.} {\bf B670} (2003) 479--507,
  [\href{http://xxx.lanl.gov/abs/math-ph/0304032}{{\tt math-ph/0304032}}].

\bibitem{Lohmayer:2008bd}
R.~Lohmayer, H.~Neuberger, and T.~Wettig, {\it {Possible large-N transitions
  for complex Wilson loop matrices}},  {\em JHEP} {\bf 11} (2008) 053,
  [\href{http://xxx.lanl.gov/abs/0810.1058}{{\tt arXiv:0810.1058}}].

\bibitem{Gross:1993hu}
D.~J. Gross and W.~Taylor, {\it {Two-dimensional QCD is a string theory}},
  {\em Nucl. Phys.} {\bf B400} (1993) 181--210,
  [\href{http://xxx.lanl.gov/abs/hep-th/9301068}{{\tt hep-th/9301068}}].

\bibitem{Dunne:1988ih}
G.~V. Dunne, {\it {Negative dimensional groups in quantum physics}},  {\em J.
  Phys.} {\bf A22} (1989) 1719.

\bibitem{Mehta:1991}
M.~L. Mehta, {\em Random Matrices}.
\newblock Academic Press, San Diego, 2nd~ed., 1991.

\bibitem{Wilf:1978}
H.~S. Wilf, {\em Mathematics for the Physical Sciences}.
\newblock Dover, 1978.

\bibitem{Szego:1991}
G.~{Szeg\"o}, {\em Orthogonal Polynomials}.
\newblock American Mathematical Society, Providence, RI, 1991.

\bibitem{Perelomov:1965ab}
A.~M. Perelomov and V.~M. Popov, {\it Casimir operators for the unitary group},
   {\em JETP Letters} {\bf 1} (1965) 160--162.

\bibitem{Senouf:1996ab}
D.~Senouf, {\it {Asymptotic and numerical approximations of the zeros of
  Fourier integrals}},  {\em SIAM J. Math. Anal.} {\bf 27} (1996) 1102--1128.

\bibitem{Blaizot:2009kj}
J.-P. Blaizot and M.~A. Nowak, {\it {Universal shocks in random matrix
  theory}},  \href{http://xxx.lanl.gov/abs/0902.2223}{{\tt arXiv:0902.2223}}.

\bibitem{Janik:2004tw}
R.~A. Janik and W.~Wieczorek, {\it {Multiplying unitary random matrices --
  universality and spectral properties}},  {\em J. Phys. A: Math. Gen.} {\bf
  37} (2004) 6521--6529.

\end{thebibliography}\endgroup

\end{document}